\PassOptionsToPackage{english}{babel}
\documentclass[aps,reprint,pra,superscriptaddress,nofootinbib]{revtex4-2}
\usepackage{dcolumn}
\usepackage[utf8]{inputenc}
\usepackage{amsmath}
\usepackage{amsfonts}
\usepackage{amsthm}
\usepackage{amssymb}
\usepackage{graphicx}
\usepackage{hyperref}
\usepackage{xcolor}
\usepackage{adjustbox}
\usepackage{comment}
\usepackage[toc,page]{appendix}
\usepackage[english,ngerman]{babel}
\usepackage{cancel}
\usepackage{xcolor}
\usepackage[normalem]{ulem}
\usepackage{tikz}
\usepackage{xcolor}
\usepackage{lipsum} 
\newtheorem{result}{Result}

\newtheorem{corollary}{Corollary}[result]

\newtheorem{remark}{Remark}[result]

\usepackage{tikz}
\usetikzlibrary{quantikz2}

\begin{document}
\selectlanguage{english}
\title{Role of non-classicality in mediated spatial quantum correlations}

\author{Salvatore Raia}
\thanks{Current address: Scuola Normale Superiore, Pisa, Italy}

\email{\\ salvatore.raia@sns.it}

\affiliation{Scuola Normale Superiore, Piazza dei Cavalieri 7, 56126 Pisa, Italy}
\affiliation{Dipartimento di Fisica, Università di Pisa, Largo Bruno Pontecorvo 3, 56127 Pisa, Italy}
\affiliation{Clarendon Laboratory, University of Oxford, Parks Road, Oxford OX1 3PU, United Kingdom}

\author{Giuseppe Di Pietra}
\affiliation{Clarendon Laboratory, University of Oxford, Parks Road, Oxford OX1 3PU, United Kingdom}

\author{Chiara Marletto}
\affiliation{Clarendon Laboratory, University of Oxford, Parks Road, Oxford OX1 3PU, United Kingdom}

\date{\today}

\allowdisplaybreaks
\begin{abstract}
The study of non-classicality is essential to understand the quantum-to-classical transition in physical systems. Recently, a witness of non-classicality has been proposed, linking the ability of a system (``the mediator") to create quantum correlations between two quantum probes with its non-classicality, intended as the existence of at least two non-commuting variables. Here, we propose a new inequality that quantitatively links the increase in quantum correlations between the probes to a function of the non-commutativity of the mediator's observables. We test the inequality for various degrees of non-classicality of the mediator, from fully quantum to fully classical. This quantum-to-classical transition is simulated via a phase-flip channel applied to the mediator, inducing an effective reduction of the non-commutativity of its variables. Our results provide a general framework for witnessing non-classicality, assessing the non-classicality of a system via its intrinsic properties, independently of the specific chosen interaction dynamics. 
\end{abstract}

\maketitle

\section{Introduction}
Understanding what makes quantum systems different from classical systems has been a central issue in quantum information and quantum foundations. One way to characterise the quantumness of a physical system is via its \textit{non-classicality}, here defined as the fact that it has at least two variables that do \textit{not} commute (i.e., that cannot be measured simultaneously to arbitrarily high accuracy by the same device) \cite{marletto_witnessing_2020}. {A system would thus be defined as classical if all its degrees of freedom can be represented by an abelian algebra, where all the operators commute with each other. This definition is in line with several models embedding classicality in a quantum framework (e.g., Koopmanian models \cite{koopman_hamiltonian_1931}), or imposing supeselection rules that effectively forbid non-commuting observables, as in \cite{sherry_interaction_1978}.}

\begin{figure}
\centering
    \includegraphics[width=0.49\columnwidth]{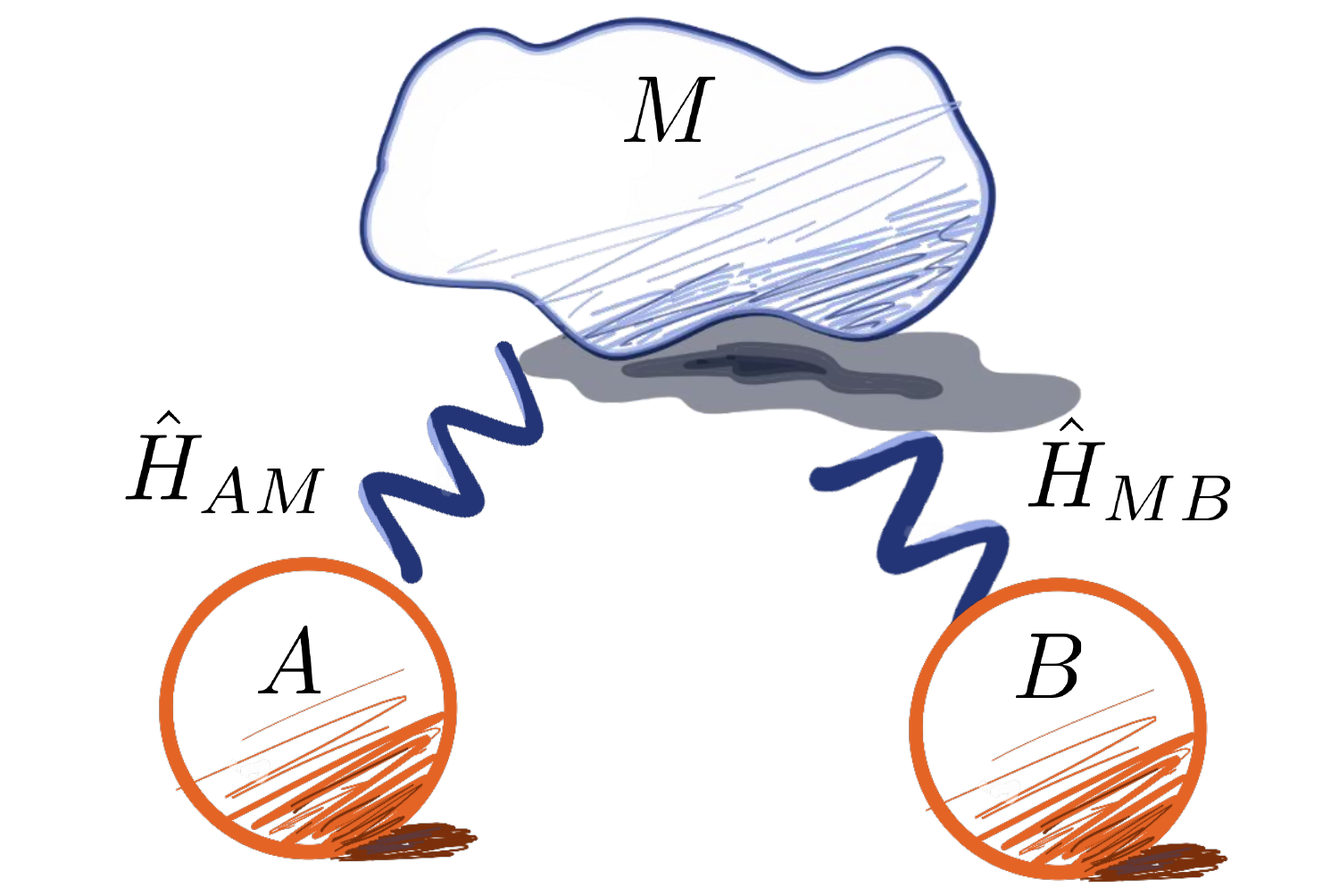}
    \label{fig:sub1}
\hfill
    \includegraphics[width=0.49\columnwidth]{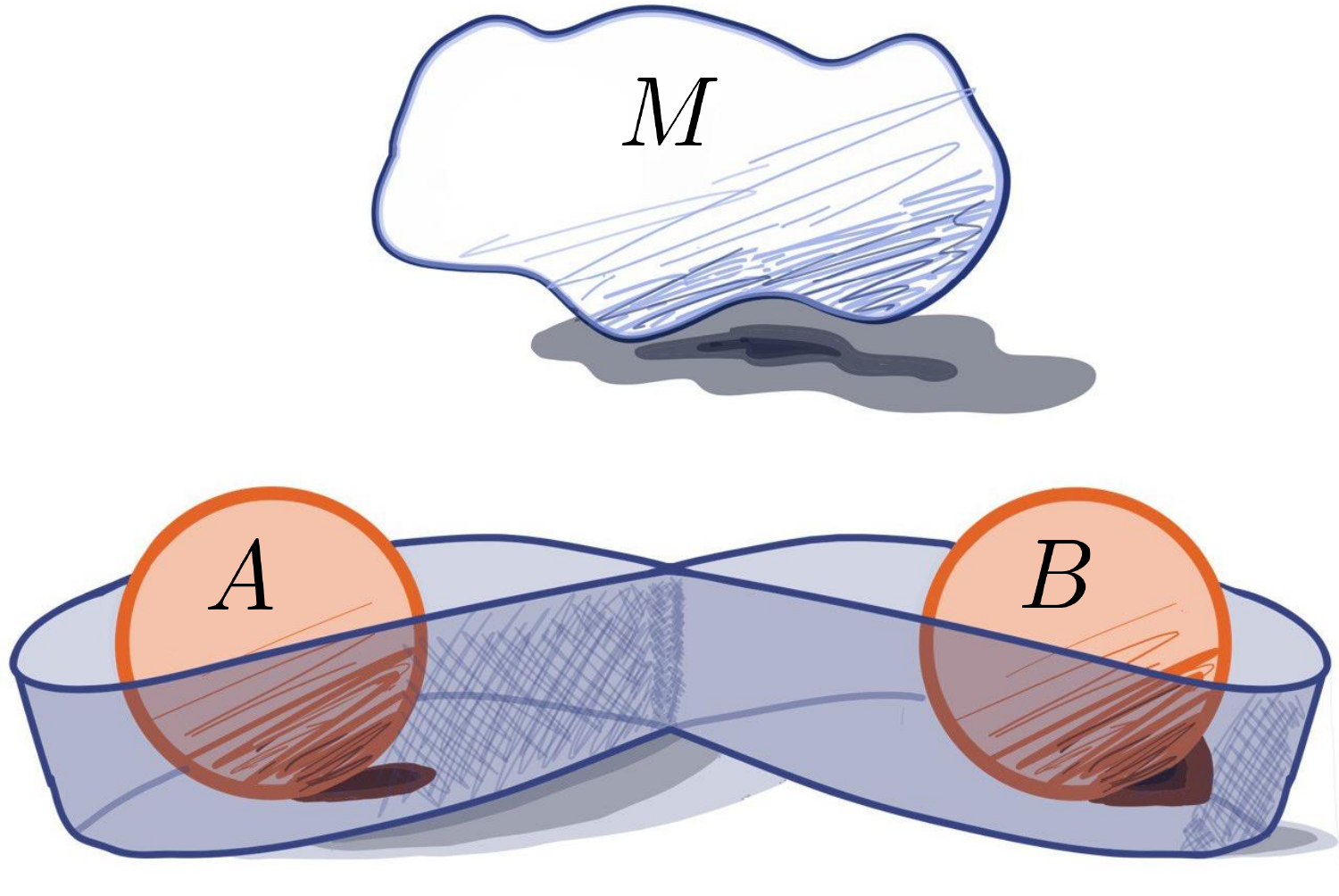}
    \label{fig:sub2}
\caption{(a): The two probes $A$ and $B$ in a separable initial state interacting via the mediator $M$ through the Hamiltonians $\hat{H}_{AM}, \hat{H}_{MB}$. (b): The two probes $A$ and $B$ are in a final entangled state after the mediated interaction occurs. Detecting the entanglement between $A$ and $B$ is then sufficient to infer the non-classicality of $M$ \cite{bose_spin_2017, PRL_Marletto_Vedral_2017}.}
\label{fig:BMV}
\end{figure}

{How could one determine whether a system is non-classical? The most obvious solution would be to directly measure the investigated system and check if two or more observables obey the uncertainty principle, one of the pillars of quantum theory \cite{heisenberg_uber_1927}.}
{However, it is not always possible to directly measure such observables, especially when facing complex systems with an untractable dynamics, like biological ones or systems for which quantum theory may break down, as in the case of gravity.
As a result, the question of if, and where, the cut between classical and quantum world should be traced is crucial for the universality of quantum theory.}
{Assuming such a cut exists, the system one should face is known as a \textit{hybrid} system, where a quantum sector $\mathcal{Q}$ made of fully fledged quantum systems (qubits, for example), interacts with a system $M$ whose quantum nature is unknown, potentially classical. In this setup, the quantum sector could be exploited to design \textit{indirect} tests of the unknown sector's non-classicality, capable of inferring the necessary quantum description of $M$ by ruling out the hypothesis that could be described by a fully classical model. Furthermore, these tests should work as \textit{sufficient} conditions for $M$'s non-classicality: by observing the effects of $M$ on $\mathcal{Q}$, one could experimentally test if $M$ must be described by the quantum formalism.}
Recently, new experiments have been proposed to indirectly witness the non-classicality of a mediator of interactions between two quantum probes \cite{marletto_quantum-information_2025}. 
These experiments suggest that the creation of entanglement between two quantum probes, which interact locally \cite{di_pietra_role_2025} only through a mediator $M$, can serve as a witness of its non-classical nature (see Figure \ref{fig:BMV}). An immediate application of this argument is a witness to the non-classicality of the gravitational field, as described in \cite{PRL_Marletto_Vedral_2017, bose_spin_2017}; however, the witness is more generally applicable \cite{marletto_witnessing_2020}. 

Interestingly, a \textit{general method} for \textit{quantitatively} assessing the witness, independent of the chosen model for the mediator and its interactions with the probes, is currently lacking. {In fact, the witness of non-classicality, as originally proposed, does not explicitly link the amount of entanglement generated by the probes to the non-classicality of the mediator \textit{independently of} the choice of the interaction Hamiltonians.} This prompts the need for a broader quantitative framework capable of encompassing a wide range of possible models for the systems and the interactions involved. Addressing this need is one of the primary motivations for this work.
 
{A significant step towards this direction has been taken. In Refs. \cite{krisnanda_detecting_2018,ganardi_quantitative_2024} the authors established a causal relation between the increase of mediated quantum correlation shared by the probes and the \textit{non-classicality of interactions}, expressed as the commutator of the interaction Hamiltonians $\hat{H}_{AM}, \hat{H}_{MB}$. 
Here, we first establish a \textit{quantitative connection} between the gain in quantum correlations of the two probes, $\Delta Q_{corr \ A:B}$, and a function of the non-classicality of interactions, specifically in the scenario where the mediator is unknown and experimentally inaccessible. Subsequently, we reformulate the non-classicality of interactions in terms of the \textit{mediator's non-classicality}, and thus express it as a function of the commutators of the mediator's observables.
When compared to the non-classicality of interactions, the non-classicality of the mediator is an intrinsic property of the system rather than specific to the probes' relation to the mediator. Additionally, the inequality is general because it is valid for any mediated interaction involving the two probes and the mediator.}

We also investigate the impact of {classicalisation} on the mediator's observables, {modelled as} a phase-flip channel applied to the mediator. We show that as the phase-flip parameter $p$ approaches $\frac{1}{2}$, the commutator value approaches zero; this simulates a transition from a fully quantum system to a fully classical system,  which{, as we prove,} is unable to create quantum correlations between the probes. 

Our result offers a deeper understanding of the mediator's role as a quantum channel capable of distributing entanglement between two communicating parties, as it focuses on an intrinsic characterisation of the mediator rather than how it dynamically interacts with the probes. 

\section{Characterisation of non-classical interactions}\label{sec:2}
Suppose our closed system is made of two quantum probes, $A$ and $B$, and a mediator $M$, all with finite-dimensional Hilbert spaces. The interactions are assumed to be \textit{local}, in the sense that there is no direct $A$--$B$ coupling \cite{di_pietra_role_2025}, and the total Hamiltonian takes the form
\begin{equation}
\hat{H}=\hat{H}_{AM}+\hat{H}_{MB},
\label{eq:total_H_mediated}
\end{equation}
where the interaction terms $\hat{H}_{AM}$ and $\hat{H}_{MB}$ are otherwise left unspecified.
{Since the dynamics of $AMB$ is assumed to be closed, the time evolution is unitary and is generated by \eqref{eq:total_H_mediated}. In particular, the global evolution operator is:
\begin{equation}
U_{AMB}(t)\vcentcolon=e^{-it(\hat H_{AM}+\hat H_{MB})},
\label{eq:U_AMB_def_main}
\end{equation}}
{
with the associated unitary channel defined as:
\begin{equation}
\Lambda_{AMB}(\rho)\vcentcolon=U_{AMB}(t)\,\rho\,U_{AMB}^\dagger(t).
\label{eq:Lambda_AMB_def_main}
\end{equation}}

{In the non-classicality witness scenario, the mediator $M$ is experimentally inaccessible; hence, to infer its possible non-classical nature, we need to test its effects through measurements on the two probes. The quantity of interest is the possible increase in quantum correlations between the probes induced by their interaction with the mediator. To retain the highest possible level of generality, a broad class of quantum correlation measures $Q$ is considered. Specifically, $Q$ is assumed to be a $gd$-continuous function, a property shared by commonly used quantum correlation measures such as entropy of entanglement and logarithmic negativity \cite{donald_continuity_1999, plenio_logarithmic_2005}. Being $gd$-continuous implies that $Q$ is monotonic under local processing and satisfies the continuity condition:
\begin{equation}
|Q(x)-Q(y)| \leq g(d(x,y)),
\end{equation}
where $g$ is an invertible monotonically increasing function with the property $\lim_{s \to 0} g(s) = 0$ and $d$ is a distance between states $x$ and $y$.
}

{We then define the increase in quantum correlation between the two quantum probes $A$ and $B$ by
\begin{equation}
\Delta Q_{corr \ A:B}
\vcentcolon=
Q_{A:B}(\rho_t) - \mathcal{B}(Q(\rho_0)) .
\label{eq:DeltaQ_def}
\end{equation}
The quantity $Q_{A:B}(\rho_t)$ is the amount of accessible (i.e.\ measurable) quantum correlation of the evolved state of the two probes' state $\rho_t$ after the interaction. The evolved state is defined as $\rho_t = \mathrm{tr}_M[\Lambda_{AMB}(\rho_{AMB}(0))]$, while $\mathcal{B}(Q(\rho_0))$ is the maximum possible initial quantum correlation between $A$ and $B$ given their initial state $\rho_0 = \mathrm{tr}_M[\rho_{AMB}]$. An exact definition of $\mathcal{B}$ is given in Appendix \ref{sec:appendixA1} (see \eqref{eq:app_B_def} and \eqref{eq:app_I_def}).}

{In \cite{ganardi_quantitative_2024, krisnanda_detecting_2018} it has been shown that if the local interactions are classical, meaning that $\hat{H}_{AM}$ and $\hat{H}_{MB}$ commute, then the amount of quantum correlation $Q$ between $A$ and $B$ after the interaction cannot exceed the maximum initial correlation, so that:
\begin{equation}
[\hat{H}_{AM},\hat{H}_{MB}]=0
\;\Longrightarrow\;
\Delta Q_{corr \ A:B}\leq 0 .
\label{eq:comm_implies_no_gain}
\end{equation}}

{Before introducing the next results of \cite{ganardi_quantitative_2024, krisnanda_detecting_2018}, we need the notion of a \textit{decomposable} map. A map that acts on the tripartite system, $\lambda_{AMB}(\rho)$, is decomposable if there exist two CPTP maps $\lambda_{AM}$ and $\lambda_{MB}$ such that:
\begin{equation}
\lambda_{AMB}(\rho)=\lambda_{MB}\circ \lambda_{AM}(\rho).
\label{eq:decomp_def}
\end{equation}
In the non-classicality witness scenario, we cannot fully determine the map induced by the local interactions \eqref{eq:total_H_mediated}, as the mediator is unknown and non-accessible. We can only access the marginal map $\Lambda_{AB}=\mathrm{tr}_M[\Lambda_{AMB}]$ given how the mediator acts on the quantum probes. The global map $\Lambda_{AMB}$ can be found among the Stinespring dilations of $\Lambda_{AB}$, with the appropriate dimension $m$. It is found in \cite{ganardi_quantitative_2024, krisnanda_detecting_2018} that if the interactions are classical, then there exists a Stinespring dilation of $\Lambda_{AB}$ of dimension $m\geq d_M)$ which is decomposable and where $d_M$ is the dimension of the mediator's Hilbert space. We call the space of all such maps' dilations $\overline{DEC}(m)$, and we can summarise this implication as:
\begin{equation}
[\hat{H}_{AM},\hat{H}_{MB}]=0
\implies
\Lambda_{AB}\in \overline{DEC}(m).
\label{eq:comm_implies_dec}
\end{equation}}

{Crucially, the non-decomposability of the induced map $\Lambda_{AB}$ admits a direct dynamical interpretation, as it implies the non-classicality of the mediated interaction:
\begin{equation}
\Lambda_{AB}\notin \overline{DEC}(m)
\iff
[\hat{H}_{AM},\hat{H}_{MB}] \neq 0 .
\label{eq:nondec_implies_noncomm}
\end{equation}}

{With these definitions in place, one of the main results of \cite{krisnanda_detecting_2018,ganardi_quantitative_2024} states that if the total quantum correlation shared between $A$ and $B$ increases, then the marginal map $\Lambda_{AB}$ is non-decomposable, that is:
\begin{equation}
\Delta Q_{corr \ A:B} > 0
\;\Longrightarrow\;
\Lambda_{AB}\notin \overline{DEC}(m).
\label{DeltaQ>0}
\end{equation}}

{We stress that $\Delta Q_{corr \ A:B}$ depends only on quantities related to the two probes $A$ and $B$, which can be measured in the laboratory. Moreover, no further assumption is made on  $\hat{H}$, $\hat{H}_{AM}$ and $\hat{H}_{MB}$ beyond locality and finite dimensionality, and the connection between the increase of quantum correlations and the non-classicality of interactions is therefore valid for any of the allowed local Hamiltonians.}
\\

Let us now proceed with our discussion by introducing one of our new results, which sets the increase in quantum correlation between the probes as a lower bound to a \textit{function} of the non-classicality of interactions. 
This result is derived under the same assumptions of the previous works \cite{krisnanda_detecting_2018,ganardi_quantitative_2024}: $A$, $B$, and $M$ are defined within finite-dimensional Hilbert spaces, and the Hamiltonians satisfy locality conditions, meaning that no direct interaction between $A$ and $B$ is allowed. Furthermore, the allowed Hamiltonians remain unspecified throughout the derivation, ensuring the generality of the result.

{This result is obtained by building on the distance-based inequality derived in \cite{ganardi_quantitative_2024}, which bounds the experimentally accessible quantity $g^{-1}\!\bigl(\Delta Q_{\mathrm{corr}\,A:B}\bigr)$ in terms of the distance between a dilation of the induced marginal map $\Lambda_{AB}$ and a suitable dilation of a decomposable map. Here, $g^{-1}$ denotes the inverse of the function $g$ appearing in the definition of $gd$-continuity of the correlation measure $Q$.
As shown in Appendix~\ref{sec:appendixA2}, this distance can be further bounded by the spectral norm of a difference of unitaries associated with the local interaction Hamiltonians, namely:}
{
\begin{equation}
\bigl\|U_{AMB}(t)-U_{AM}(t)\,U_{MB}(t)\bigr\|_\infty,
\end{equation}
where $U_{AMB}(t)$ is defined in \eqref{eq:U_AMB_def_main}, while $U_{AM}(t)$ and $U_{MB}(t)$ are defined as
\begin{equation}
U_{AM}(t)\vcentcolon=e^{-it\hat H_{AM}},
\qquad
U_{MB}(t)\vcentcolon=e^{-it\hat H_{MB}}.
\label{eq:local_UAM_UMB_def}
\end{equation}}
{This quantity measures, in operator norm, the deviation of the true global evolution of $AMB$ from the decomposed evolution obtained by concatenating the evolutions generated by $\hat H_{AM}$ and $\hat H_{MB}$.}

{To make the dependence on non-commutativity explicit, we consider the Zassenhaus factorisation of the global unitary,
\begin{equation}
U_{AMB}(t)
=
U_{AM}(t)\,U_{MB}(t)\,
\prod_{n=2}^{\infty} \exp\!\bigl[(-it)^n Z_n\bigr],
\label{eq:U_AMB_Zassenhaus_def_main}
\end{equation}
and define the associated Zassenhaus correction unitary
\begin{equation}
U_C(t)\vcentcolon=
\prod_{n=2}^{\infty} \exp\!\bigl[(-it)^n Z_n\bigr]
=
e^{-it\hat H_C},
\label{eq:Ualph_Halp_def_main}
\end{equation}
where $\hat H_C$ is the corresponding Hermitian generator of $U_C(t)$.
Appendix~\ref{sec:appendixA2} shows that:
\begin{equation}
\bigl\|U_{AMB}(t)-U_{AM}(t)U_{MB}(t)\bigr\|_\infty
=
\bigl\|U_C(t)-\mathbb{I}\bigr\|_\infty,
\end{equation}
and that the latter admits a closed functional form in terms of $\hat H_C$.}

{We now present the outcome of this derivation, a quantitative relation that links $g^{-1}\!\bigl(\Delta Q_{\mathrm{corr}\,A:B}\bigr)$ directly to a function of the non-classicality of interactions.}

{\begin{result}
Consider a closed finite-dimensional tripartite system {$A\oplus M\oplus B$} with mediated Hamiltonian $\hat H=\hat H_{AM}+\hat H_{MB}$. For any $gd$-continuous quantum correlation measure $Q$, for the increase in quantum correlation $\Delta Q_{\mathrm{corr}\,A:B}$ defined in \eqref{eq:DeltaQ_def} and the Zassenhaus correction Hamiltonian $\hat{H}_C$ defined in \eqref{eq:Ualph_Halp_def_main}, one has:
\label{result1}
\end{result}}
{\begin{equation}
g^{-1}\!\bigl(\Delta Q_{\mathrm{corr}\,A:B}\bigr)\;\le\;2\;\Bigl\|\sin\!\Bigl(\frac{t}{2}\,\hat H_C\, \Bigr)\Bigr\|_\infty \quad\forall t.
\label{main_res_explicit}
\end{equation}}

{As shown in Appendix~\ref{sec:appendixA1}, $\hat H_C$ is completely determined by the interaction Hamiltonians $\hat H_{AM}$ and $\hat H_{MB}$, and it vanishes whenever they commute, i.e.\ $[\hat H_{AM},\hat H_{MB}]=0 \Rightarrow \hat H_C=0$.}
{In the short-time regime, a second-order expansion in $t$, detailed in Appendix~\ref{sec:appendixA2}, yields an especially transparent form of the bound, since it makes the dependence on the commutator $[\hat H_{AM},\hat H_{MB}]$ fully explicit:}
{
\begin{equation}
g^{-1}\!\bigl(\Delta Q_{\mathrm{corr}\,A:B}\bigr)
\leq
2\,\Bigl\|\sin\!\Bigl(\frac{i t^2}{4}\,[\hat{H}_{AM}, \hat{H}_{MB}]\Bigr)\Bigr\|_\infty.
\label{main_res_explicit_[]}
\end{equation}}
{This inequality relates the gain in quantum correlations between $A$ and $B$ to a function of the commutator $[\hat{H}_{AM}, \hat{H}_{MB}]$.
Equivalently, by the same expansion, the leading nontrivial order in $t$ implies that the generator entering \eqref{main_res_explicit} is:}
{\begin{equation}
\hat{H}_C \approx \frac{i t}{2}\,[\hat{H}_{AM}, \hat{H}_{MB}].
\label{eq:Halpha_Z2_approx}
\end{equation}}
{For later convenience, we introduce a compact notation for the right-hand side of the short-time bound. Define the map
\begin{equation}
f(H)\vcentcolon=
g\!\left(2\left\|\sin\!\left(\frac{i t^2}{4}\,H\right)\right\|_\infty\right),
\label{eq:def_f_commutator}
\end{equation}
so that \eqref{main_res_explicit_[]} can be equivalently written as
\begin{equation}
\Delta Q_{\mathrm{corr}\,A:B}\le  f\!\bigl([\hat H_{AM},\hat H_{MB}]\bigr).
\label{eq:DeltaQ_le_f_commutator}
\end{equation}
This definition is purely notational; once the correlation measure $Q$ (hence $g$) and the interaction time $t$ are fixed, the bound depends on the dynamics only through the commutator $[\hat H_{AM},\hat H_{MB}]$.}

{Taken together, inequalities \eqref{main_res_explicit} and \eqref{main_res_explicit_[]} establish a quantitative link between the gain in quantum correlations of the two probes and the non-classicality of the underlying interactions, in the setting where the mediator is unknown and experimentally inaccessible. This constitutes the first main result of this work.}
\\

{Before proceeding further, it is worth making a few remarks on the main inequality
\eqref{main_res_explicit}. If the interactions are classical, namely if $[\hat H_{AM},\hat H_{MB}]=0$, then, Appendix~\ref{sec:appendixA1} shows that the Zassenhaus correction is trivial, $U_C(t)=\mathbb{I}$, and hence $\hat H_C=0$ (see \eqref{eq:commuting_implies_Ualpha_trivial}).
Consequently, since $\sin\!\bigl(\tfrac{t}{2}\hat H_C\bigr)=0$, the right-hand side of \eqref{main_res_explicit} vanishes, and we obtain $g^{-1}\!\bigl(\Delta Q_{\mathrm{corr}\,A:B}\bigr)\le 0.$
Since $g$ is monotonically increasing, invertible, and satisfies $\lim_{s\to 0} g(s)=0$, it follows that:
\begin{equation}
\Delta Q_{\mathrm{corr}\,A:B}\le 0.
\end{equation}
In particular, the condition \eqref{DeltaQ>0} cannot be satisfied. In agreement with \cite{ganardi_quantitative_2024}, we summarise this consequence as follows.}

{\begin{corollary}\label{cor:classical_no_gain}
\textit{Classical interactions cannot increase quantum correlations.}
If $[\hat H_{AM},\hat H_{MB}]=0$, then
$\Delta Q_{\mathrm{corr}\,A:B}\le 0$.
\end{corollary}}

{Conversely, if an experiment reveals an increase in quantum correlations, namely $\Delta Q_{\mathrm{corr}\,A:B}>0$, then, by the properties of $g$ recalled above,
$g^{-1}\!\bigl(\Delta Q_{\mathrm{corr}\,A:B}\bigr)>0$.
By \eqref{main_res_explicit}, this can only occur if the right-hand side is strictly positive, hence $\hat H_C\neq 0$.
Using \eqref{eq:Halpha_nonzero_implies_noncommuting} in Appendix~\ref{sec:appendixA1}, we conclude that the interaction Hamiltonians cannot commute, i.e.\ $[\hat H_{AM},\hat H_{MB}]\neq 0$.
This is consistent with the explicit second-order form \eqref{main_res_explicit_[]}, where the leading nontrivial contribution
is governed by the commutator $[\hat H_{AM},\hat H_{MB}]$.}
{\begin{corollary}\label{cor:positive_gain_nonclassical_interactions}
\textit{Increase in quantum correlation rules out classical interactions.}
If $\Delta Q_{\mathrm{corr}\,A:B}>0,$ then necessarily $[\hat H_{AM},\hat H_{MB}]\neq 0,$ equivalently, $\hat H_C\neq 0$.
\end{corollary}}

{A further remark concerns the size and achievability of the right-hand side of \eqref{main_res_explicit}.
Since $\|\sin(X)\|_\infty\le 1$ for any Hermitian operator $X$, the bound is limited as
\begin{equation}
2\Bigl\|\sin\!\Bigl(\frac{t}{2}\hat H_C\Bigr)\Bigr\|_\infty \le 2,
\end{equation}
so that its maximum is universal, i.e.\ independent of the specific choice of $\hat H_{AM}$ and $\hat H_{MB}$.}

{Moreover, for any fixed $\hat H_C\neq 0$, this maximal value can be attained by a
suitable choice of the interaction time $t$.
As shown in Appendix~\ref{sec:appendixA2} (see Eq.~\eqref{eq:tstar_def}), the earliest time at which this occurs is
controlled by the largest eigenvalue of $\hat H_C$, and scales as
$t_\ast=\pi/\max_j|\lambda_j|$, where $\{\lambda_j\}$ are the eigenvalues of $\hat H_C$.
In this sense, while the maximal value of the bound is universal, the shortest interaction
time needed to approach it depends on the spectral structure of $\hat H_C$.}

{Finally, in the short-time regime where
$\hat H_C \approx \tfrac{i t}{2}[\hat H_{AM},\hat H_{MB}]$ (see \eqref{eq:app_Halpha_Z2_approx} in Appendix~\ref{sec:appendixA2}), it is natural to expect that larger non-commutativity, as
quantified for instance by $\|[\hat H_{AM},\hat H_{MB}]\|_\infty$, tends at leading order to increase the initial growth rate of the right-hand side of the bound \eqref{main_res_explicit_[]} as a function of $t$, and may therefore allow the maximum to be approached more rapidly. A systematic characterisation of these time scales is left for future work.}

\section{From non-classical interactions to non-classical mediator}\label{sec:3}
The goal of the following section will be to generally express the non-classicality of interactions, i.e. non-commutativity of interaction Hamiltonians, the quantity appearing in \eqref{main_res_explicit_[]}, in terms of the \textit{non-classicality of the mediator}, namely the non-commutativity of the mediator's observables, shifting from a model-dependent point of view to a model-independent one.

The starting point is to define more precisely the most general $\hat{H}_{AM}, \hat{H}_{MB}$ and then explicitly calculate their commutator, {to show that it ultimately depends on the commutator of the mediator's observables. }
{This shows that the ability of a mediator to create quantum correlations between two spatially separated quantum systems is due to its non-classicality and not to the way it interacts with these probes.} 

{We shall start by defining the quantum probes $A$, $B$ and the mediator $M$. }
Following the assumptions of finite dimensions, we have two possible very general choices for the Hilbert spaces $\mathcal{H}_A,  \mathcal{H}_M, \mathcal{H}_B$. The first is to consider a generic d-dimensional Hilbert space, 
while the second possibility is to choose a tensor product of qubit Hilbert spaces $\mathcal{H}_2 \otimes \ldots \otimes \mathcal{H}_2$, that, given the universality of quantum computation, would allow us to use a certain amount of qubits to simulate any quantum system of finite dimension \cite{deutsch_universality_1995}.
To simplify the notation and without loss of generality, we set the mediator $M$ to be a chain of $T$ qubits with Hilbert space dimension $d_M = 2^{T}$, while the probes $A$ and $B$ to be $d_A, d_B$ dimensional systems described by Hilbert spaces with $SU(N)$ operator algebra. Thus, the most general $\hat{H}_{AM}, \hat{H}_{MB}$ Hamiltonians will be: 
\begin{equation}
\begin{split}
&\hat{H}_{AM} = \\
&\sum_{i, j_1, j_2, \ldots j_{T}}\alpha_{i, j_1, j_2, \ldots j_{T}} (a_i \otimes m_{j_1}\otimes m_{j_2} \otimes \ldots \otimes  m_{j_{T}} \otimes \mathbb{I}_{B}),\\
&  \\
&\hat{H}_{MB} =\\
&\sum_{l_1, l_2, \ldots , l_{T}, k}\beta_{l_1, l_2, \ldots , l_{T}, k} (\mathbb{I}_{A} \otimes m_{l_1}\otimes m_{l_2} \otimes \ldots \otimes  m_{l_{T}} \otimes b_k),
\label{eq:most_general_HAM_HMB}
\end{split}
\end{equation}
where $\alpha$ and $\beta$ are real-valued coefficients, $a_i$ and $b_k$ are $SU(d_A)$ and $SU(d_B)$ operators including the identity; $\{ m_{j_1}\}_{j_1 = 0,\ldots,3} = \{\mathbb{I},X,Y,Z\}$ is the set of the Pauli operators plus the identity for the first qubit of the mediator in $\hat{H}_{AM}$, where the number used as a subscript indicates which qubit of the mediator the operator refers to. To distinguish the mediator's operators appearing in $\hat{H}_{MB}$ from those appearing in $\hat{H}_{AM}$, the subscript $l$ replaces $j$.
Most importantly, the identity operators appearing in the tensor products come from imposing the \textit{locality} condition on the interactions. This avoids any direct interaction term of the pair $AB$, thus enforcing locality. 

{Considering a mediator $M$ composed of $T$ qubits, the commutator of the interaction Hamiltonians admits the explicit expansion:}
\begin{equation}
\begin{split}
&[\hat{H}_{AM}, \hat{H}_{MB}] =\\
& = \sum_{i, j_1, j_2, \ldots j_{T}, l_1, l_2, \ldots , l_{T}, k} \alpha_{i, j_1, j_2, \ldots j_{T}} \beta_{l_1, l_2, \ldots , l_{T}, k} \quad a_i \otimes\\
& \quad \otimes \bigg( [m_{j_1}, m_{l_1}] \otimes m_{j_2}m_{l_2} \otimes \ldots \otimes  m_{j_{T}} m_{l_{T}}  \\
& \quad  \quad + \, m_{l_1} m_{j_1} \otimes [m_{j_2}, m_{l_2}] \otimes \ldots \otimes  m_{j_{T}} m_{l_{T}} \\
& \quad \quad + \, m_{l_1} m_{j_1} \otimes m_{l_2} m_{j_2} \otimes \ldots \otimes  [m_{j_{T}}, m_{l_{T}}]\bigg) \otimes b_k.
\label{eq:nonclassmediator}
\end{split}
\end{equation}
{
Equation \eqref{eq:nonclassmediator} constitutes one of the main results of this work. Its full derivation is reported in Appendix~\ref{sec:appendixA3}. Most importantly, \eqref{eq:nonclassmediator} shows that $[\hat H_{AM},\hat H_{MB}]$ can be non-vanishing only if at least one commutator of mediator observables appearing in the two couplings is nonzero, i.e.\ only if there exists at least one pair such that $[m_{j_r},m_{l_r}]\neq 0$ for some mediator-qubit label $r$. In addition, Appendix~\ref{sec:appendixA3} provides a dedicated analysis of the two-qubit case, identifying which combinations of mediator observables can lead to a non-vanishing Hamiltonian commutator. We summarise this implication as follows.}
{
\begin{result}
\label{res:nonclassical_interactions_imply_nonclassical_mediator}
\textit{Non-classical interactions imply a non-classical mediator.}
Consider a mediated Hamiltonian $\hat H=\hat H_{AM}+\hat H_{MB}$ with $\hat H_{AM},\hat H_{MB}$ of the form \eqref{eq:most_general_HAM_HMB}, then their commutator is given by \eqref{eq:nonclassmediator}.
If $[\hat H_{AM},\hat H_{MB}] \neq 0$, then there exists at least one pair of mediator observables appearing in the two Hamiltonians whose commutator is non-vanishing. In particular, in the expansion \eqref{eq:nonclassmediator}, at least one mediator commutator satisfies \([m_{j_r},m_{l_r}] \neq 0\)
for some mediator-qubit label $r$.
\end{result}}
\noindent
Equivalently, if all mediator observables that appear in the interaction terms commute with each other, then the interaction Hamiltonians commute. This gives the following corollary.

{
\begin{corollary}\label{cor:classical_mediator_implies_classical_interactions}
\textit{A classical mediator implies classical interactions.}
If, for all mediator operators appearing in \eqref{eq:most_general_HAM_HMB}, one has
$[m_{j_r},m_{l_r}]=0$ for every mediator-qubit label $r$, then
$[\hat{H}_{AM}, \hat{H}_{MB}] = 0$.
\end{corollary}}
{
\noindent
Importantly, the converse does not generally hold: a non-classical mediator does not force
$[\hat{H}_{AM},\hat{H}_{MB}]\neq 0$, because the Hamiltonians may include only a commuting subset of
mediator observables (or admit cancellations). To explain the logic of the implication, let us provide a simple example.
Let $M$ be a single qubit, so that it supports non-commuting observables (e.g.\ $[X,Z]\neq 0$).
Choose local couplings involving only the same commuting mediator observable $Z$:
\begin{equation}
\hat{H}_{AM}= a_{i} \otimes Z \otimes \mathbb{I}_B,
\qquad
\hat{H}_{MB}= \mathbb{I}_A \otimes Z \otimes b_{k},
\end{equation}
with generic Hermitian operators $a_{i}$ and $b_{k}$ acting on the two probes.
Then
\begin{equation}
[\hat{H}_{AM},\hat{H}_{MB}] = 0,
\end{equation}
even though the mediator is non-classical.
}
{
\begin{remark}\label{rem:nonclassical_mediator_commuting_interactions_example}
Mediator non-classicality is necessary but not sufficient to guarantee
non-classical interactions.
\end{remark}
}

\noindent

{
To identify when the commutator \eqref{eq:nonclassmediator} can be non-vanishing, it is useful to analyse the mediator contribution explicitly in the two-qubit case. The detailed derivation is given in Appendix \ref{sec:appendixA3}; here we summarise the outcome. 
Depending on whether the mediator indices satisfy $l_1\neq j_1$ or $l_1=j_1$, the non-vanishing terms in the mediator factor in \eqref{eq:nonclassmediator} are controlled, respectively, by the commutator
of the first-qubit operators $[m_{j_1},m_{l_1}]$ or by the commutator of the second-qubit operators
$[m_{j_2},m_{l_2}]$. In particular, in this two-qubit case, $[\hat H_{AM},\hat H_{MB}]\neq 0$ can occur only if at least a pair of non-commuting mediators' operators appear in the two Hamiltonians $\hat H_{AM}$ and $\hat H_{MB}$.}

{In general, for the specific chosen Hamiltonians, the presence of a single commutator for one of the mediator's qubits is sufficient to ensure that the commutator of the Hamiltonians is non-vanishing. This motivates the following remark.}

{\begin{remark}\label{rem:single_quantum_component}
{A \textit{single} quantum system within a composite classical-quantum mediator $M$ can be enough to mediate quantum correlations between the probes $A$ and $B$.
}
\end{remark}}

{
We now make the functional dependence of the short-time quantitative bound \eqref{main_res_explicit_[]} explicit by rewriting its right-hand side using the expansion \eqref{eq:nonclassmediator} of $[\hat H_{AM},\hat H_{MB}]$ for a mediator $M$ composed of $T$ qubits.}

{
Recalling here for convenience the bound \eqref{eq:DeltaQ_le_f_commutator} in terms of the map $f$ introduced in \eqref{eq:def_f_commutator}, we have $\Delta Q_{\mathrm{corr}\,A:B}\le f\!\bigl([\hat H_{AM},\hat H_{MB}]\bigr).$
The functional form of $f$ is fixed once the correlation measure $Q$ (hence $g$) and the interaction time $t$ are fixed, and \eqref{eq:DeltaQ_le_f_commutator} holds for \emph{any} choice of local Hamiltonians $\hat H_{AM}$ and $\hat H_{MB}$.}

{
Indeed, the coefficients $\alpha$ and $\beta$ defining the Hamiltonians in \eqref{eq:most_general_HAM_HMB} select which tensor-product factors of probe and mediator observables are activated in the interaction, and with what weights.
Crucially, the explicit form of the Hamiltonians' commutator in Eq.~\eqref{eq:nonclassmediator} shows that each potentially non-vanishing contribution contains at least one commutator of mediator observables.
Therefore, for any fixed choice of $\hat H_{AM}$ and $\hat H_{MB}$, the right-hand side of \eqref{eq:DeltaQ_le_f_commutator} depends on the mediator only through the collection of commutators of those mediator observables that appear in the two interaction terms.
For notational convenience, we denote this dependence by
\begin{equation}
f\!\bigl([\hat H_{AM},\hat H_{MB}]\bigr)\;\equiv\; f_{H}\!\bigl([m_{AM},m_{MB}]\bigr),
\label{eq:def_fH_mcomm}
\end{equation}
where $[m_{AM},m_{MB}]$ denotes the collection $\{[m_{j_r},m_{l_r}]\}_{r=1}^{T}$ of mediator commutators appearing in the expansion of $[\hat H_{AM},\hat H_{MB}]$, and the subscript $H$ reminds that the explicit expression of $f_H([m_{AM},m_{MB}])$ is obtained only after specifying the Hamiltonians (i.e.\ the coefficients $\alpha$ and $\beta$), while its functional form is universally fixed by \eqref{eq:def_f_commutator} and by the general form of the Hamiltonians' commutator \eqref{eq:nonclassmediator}.
Accordingly, we can rewrite \eqref{eq:DeltaQ_le_f_commutator} in the new form
\begin{equation}
\Delta Q_{\mathrm{corr}\,A:B}\le f_{H}\!\bigl([m_{AM},m_{MB}]\bigr).
\label{eq:DeltaQ<fH_[m,m]}
\end{equation}
In this sense, once $Q$ and $t$ are fixed, an experimentally measured gain $\Delta Q_{\mathrm{corr}\,A:B}$ is upper-bounded by a function controlled by the non-commutativity of the mediator observables, which are dynamically involved in the interaction.}

In what follows, we shall explore what happens when we send each of the commutators $ [m_{j_r},m_{l_r}]$ to 0. This is equivalent to considering a progressively more classical mediator, and formally restricting the upper bound to the possible value of $\Delta Q_{corr \ A:B}$.

\section{A classical mediator cannot create quantum correlations}\label{sec:4}
In real experimental scenarios, physical systems under investigation cannot be perfectly protected from interactions with the environment, leading to some errors and noise that inevitably modify our system by making it \textit{classical}, {and thus, with only commuting observables.}
It is thus crucial to understand the effects of this process on the capability of the unknown system to mediate quantum coherence. The results presented in this work are perfect for the scope, as \eqref{eq:DeltaQ<fH_[m,m]} shows that the upper bound to the probes' quantum correlation gain mediated by $M$, is indeed a function of its non-classicality. 
{For, we shall consider here a model that simulates a system with a lower non-classicality compared to the full quantum case. This is reflected on the commutativity of $M$'s variables, thus ultimately on $M$'s non-classicality according to the definition adopted in this work.}

A common noisy channel that might be considered for this purpose is the \textit{phase-flip channel}, which can act on our qubits, making the commutator's value of their observables vanish. This happens when the value of $p$, a free parameter of the phase flip, is $\frac{1}{2}$. This possibility is interesting {because, as we shall see,} it sets the upper bound of the inequality \eqref{eq:DeltaQ<fH_[m,m]} to 0, effectively preventing the mediator from increasing the quantum correlation across the two probes. This result suggests that a completely decohered quantum channel cannot increase the quantum correlations between two probes, as one can physically expect.

Given these considerations, {we shall assume a noise model whose overall effect is to suppress the non-commutativity of the mediator’s observables progressively, essentially making them more classical under our definition. In particular, we} allow the qubits of the mediator in our system to be affected separately by independent phase-flip channels. {This operation can be viewed as inducing a permanent modification of the mediator’s effective Hilbert space structure and of the interaction Hamiltonians $H_{AM}$ and $H_{MB}$; however, providing a physical mechanism that gives rise to such a modification lies beyond the scope of the present work and is left for future investigation. {Note that the phase-flip channel considered here is applied \textit{before} the local interactions between $A$ and $M$, and $M$ and $B$ occur, acting as a preparation of the mediator.}} 

{To formalise this description,} we shall resort to the Heisenberg picture of quantum mechanics, using the descriptors formalism \cite{deutsch_information_2000,bedard_abc_2021}. Since we have defined a base of observables, it will be enough to see how they are affected by the phase flip. 

Let \(q_x\) denote an operator representing a tensor product of the $X$ component of a qubit in our system with the identity operators on all other subsystems, for example, $q_{x_{j_1}} = \mathbb{I}_{A} \otimes X_{M_1} \otimes \mathbb{I}_{M_2} \otimes \mathbb{I}_{M_3} \otimes \dots \otimes \mathbb{I}_{B}$ where the subscript $j_1$ in $q_x$ indicates the first qubit of the mediator coupled with the probe $A$ in the Hamiltonian $\hat{H}_{AM}$. Then we have also \(q_z\), \(q_y\), with \(q_x^2 = q_y^2 = q_z^2 = I\) and the products \(q_z q_x = i q_y\), \(q_x q_z = -i q_y\) which define the algebraic properties. If a gate \(U(t_n)\) operates between time \(t_n\) and \(t_{n+1}\), we shall denote: $\mathcal{O}_x (t_{n+1}) = U(t_n)^{\dagger} \mathcal{O}_x (t_n) U(t_n)$ the operator representing the observable \(\mathcal{O}_x\) after its action. The initial conditions are fixed by choosing particular values for \(q_{x} (t_0)\), \(q_{y} (t_0)\), \(q_{z} (t_0)\), and the Heisenberg state \(\rho_H\). Using the algebraic relations for the observables defined above, the state of a qubit at time \(t_n\) is then completely specified by giving at least two components, e.g. \(\{q_{x} (t), q_{z} (t)\}\) \cite{bhole_witnesses_2020}. 

Let's recall how a phase-flip acts on a qubit and its observables. For a general observable \(\mathcal{O}_x (t)\), the phase-flip channel acts 
\begin{equation}
\hat{E} (\mathcal{O}_x (t)) = \sum_a M_a (t)^{\dagger} \mathcal{O}_x (t) M_a (t) 
\label{eq:Kraus_dephasing}
\end{equation}
where the \(M_a\) are the Kraus operators of the phase-flip channel: \(M_0 (t) = \sqrt{p}I\), \(M_1 (t) = \sqrt{1-p} \quad q_z (t)\). After this operation, the generators of the algebra of the qubit undergoing decoherence, are affected in the following way \cite{bhole_witnesses_2020}:
\begin{equation}
\begin{cases}
    \hat{E} (q_x (t)) = (2p -1) q_x (t) \\
    \hat{E} (q_y (t)) = (2p -1) q_y (t) \\
    \hat{E} (q_z (t)) = q_z (t)
\end{cases}
\end{equation}

We can now move to the Hamiltonians and apply a phase-flip \( \hat{E} \) to both \( \hat{H}_{AM} \) and \( \hat{H}_{AB} \), one per mediator's qubit in \( M \).
After the calculations provided in Appendix \ref{sec:appendixA4}, the Hamiltonians read as
\begin{equation}\begin{split}
\hat{H}_{AM}&= \hat{E}_T ( \dots \hat{E}_2 ( \hat{E}_1 ( \hat{H}_{AM} ) ) ) = \\
&= \sum_{a_i, j_1, \ldots, j_T} \alpha_{a_i, j_1, \ldots, j_T}^{'} (q_{a_i} q_{j_1} \cdot \ldots \cdot q_{j_T} q_{\mathbb{I}_{B}})
\end{split}\end{equation}
Here, each descriptor is an observable in the $AMB$ Hilbert space, so the tensor product is no longer required. As before, the qubit of the $B$ probe is set to be an identity operator, meaning that the descriptor $q_{\mathbb{I}_{B}}$ is simply a tensor product of identity operators of all the qubits in $A$, $M$ and $B$.
Similarly for \(\hat{H}_{MB} \):
\begin{equation}
\begin{split}
\hat{H}_{MB}&= \hat{E}_T ( \dots \hat{E}_2 ( \hat{E}_1 (\hat{H}_{MB} ) ) ) = \\
&= \sum_{l_1, \ldots, l_T, b_k} \beta_{l_1, \ldots, l_T, b_k}^{'} (q_{\mathbb{I}_{A}} q_{l_1} \cdot \ldots \cdot q_{l_T}  q_{b_k})
\end{split}
\label{Degraded_H_MB}
\end{equation}
These resulting Hamiltonians account for the multiple phase flips that have been applied, by storing an extra $(2p-1)$ factor in the primed coefficients $\alpha^{'}$ and $\beta^{'}$ each time an $X$ or $Y$ observable is considered for a qubit mediator (see the discussion around  \eqref{eq:app_degraded_ge_hamiltonian_HAM} in Appendix \ref{sec:appendixA4}). We now calculate the commutator for the new degraded Hamiltonians by simply following the same steps as the previous commutator in \eqref{eq:nonclassmediator} as explained in Appendix \ref{sec:appendixA4}, obtaining
\begin{equation}
\begin{split}
&[\hat{H}_{AM}, \hat{H}_{MB}] =\\
& = \sum_{a_i, j_1, j_2, \ldots j_{T}, l_1, l_2, \ldots , l_{T}, b_k} \alpha_{a_i, j_1, j_2, \ldots j_{T}}^{'} \beta_{l_1, l_2, \ldots , l_{T}, b_k}^{'} \quad q_{a_i} \\
&  \quad \bigg( [q_{j_1}, q_{l_1}]  q_{j_2}q_{l_2} \cdot \ldots \cdot q_{j_{T}} q_{l_{T}} + q_{l_1} q_{j_1} [q_{j_2}, q_{l_2}]  \cdot \ldots \cdot  q_{j_{T}} q_{l_{T}} + \\
& \quad +  q_{l_1} q_{j_1} q_{l_2} q_{j_2} \cdot \ldots \cdot   [q_{j_{T}}, q_{l_{T}}]\bigg)  q_{b_k}
\label{Degraded_Hamiltonian_Commutator}
\end{split}
\end{equation}

{
This expression has the same algebraic structure as the commutator in \eqref{eq:nonclassmediator}, with the only difference that the coefficients
$\alpha$ and $\beta$ are replaced by the phase-flip--degraded coefficients $\alpha'$ and $\beta'$, which incorporate factors of $(2p-1)$ whenever an
$X$ or $Y$ mediator observable appears. In particular, the dependence of the correlation-gain bound on decoherence enters through the degraded
Hamiltonian commutator in \eqref{Degraded_Hamiltonian_Commutator}. In direct analogy with \eqref{eq:DeltaQ<fH_[m,m]}, we therefore write
\begin{equation}
\Delta Q_{\mathrm{corr}\,A:B}\le f_{H'}\bigl([m_{AM},m_{MB}],p\bigr),
\label{eq:DeltaQ_le_fHprime_mcomm_p}
\end{equation}
where $f_{H'}([m_{AM},m_{MB}],p)$ denotes the right-hand side of the bound obtained after specifying the interaction Hamiltonians
(i.e.\ the coefficients $\alpha,\beta$) and applying the phase-flip channel with parameter $p$ to the mediator observables, so that the mediator
commutators in the expansion are rescaled by the corresponding $(2p-1)$ factors encoded in $\alpha'$ and $\beta'$.}

{As discussed in the previous section, only certain combinations of mediator observables appearing in the Hamiltonians can contribute to a non-vanishing commutator. Crucially, any potentially non-vanishing term in \eqref{Degraded_Hamiltonian_Commutator} contains at least one commutator of mediator observables, which always involves at least a $X$ or $Y$ component; thus, it is accompanied by one or two factors of $(2p-1)$.
As a consequence, the degraded commutator vanishes at $p=\frac{1}{2}$, so that $f_{H'}([m_{AM},m_{MB}],\tfrac{1}{2})=0$ and
\begin{equation}
\Delta Q_{\mathrm{corr}\,A:B}\le 0.
\end{equation}
Therefore, if an experiment observes $\Delta Q_{\mathrm{corr}\,A:B}>0$ under the assumptions of the witness, the mediator cannot be fully {classicalised} in
this phase-flip sense, and must retain non-commuting observables dynamically involved in the interaction.}

{More generally, different noise models acting on $M$ would modify the effective Hamiltonian commutator depending on the way they affect the mediator and its interaction with the probes. A detailed study of these decoherence channels {is left} for future work.}

\section{Discussion}

{
In this work, we have proposed a novel inequality quantitatively connecting the non-classicality of a mediator of interactions with its ability to create quantum correlations between two probes.
In particular, we considered a closed finite-dimensional tripartite system $AMB$ evolving under an interaction Hamiltonian $\hat H=\hat H_{AM}+\hat H_{MB}$, where $M$ mediates the interaction between the probes $A$ and $B$, see Fig. \ref{fig:BMV}. 
We have taken as our starting point the results in \cite{krisnanda_detecting_2018,ganardi_quantitative_2024}, which relate an increase of quantum correlations $\Delta Q_{\mathrm{corr}\,A:B}$ to the non-classicality of the interactions, expressed as the non-commutativity of $\hat H_{AM}$ and $\hat H_{MB}$.}

{
Our first main contribution is Result~\ref{result1}, which provides a quantitative upper bound on the experimentally accessible increase in quantum correlations $\Delta Q_{\mathrm{corr}\,A:B}$, defined in \eqref{eq:DeltaQ_def}, in terms of a closed function of the mediated dynamics, see \eqref{main_res_explicit}.
In the short-time regime, this bound admits the explicit commutator form \eqref{main_res_explicit_[]}, making the dependence on the non-commutativity of the interaction Hamiltonians transparent.
Equivalently, the short-time bound can be written in the more compact form $\Delta Q_{\mathrm{corr}\,A:B}\le f\!\bigl([\hat H_{AM},\hat H_{MB}]\bigr)$, as already given in \eqref{eq:DeltaQ_le_f_commutator}, where $f$ is defined in \eqref{eq:def_f_commutator} and is fixed by the chosen correlation measure $Q$ and by the interaction time $t$, while the dynamics enters only through the Hamiltonian commutator.
Altogether, under the sole assumptions of locality and finite dimensionality, Result~\ref{result1} yields a quantitative link between the gain in quantum correlations of the probes and the non-classicality of the interactions{, as clearly remarked by the Corollary~\ref{cor:positive_gain_nonclassical_interactions}.}}

{
Our second main contribution consists of Eq.~\eqref{eq:nonclassmediator} and Result~\ref{res:nonclassical_interactions_imply_nonclassical_mediator}, which reformulate the non-classicality of interactions in terms of the mediator's non-classicality.
Combining \eqref{eq:nonclassmediator} with the short-time bound in the compact form \eqref{eq:DeltaQ_le_f_commutator} yields \eqref{eq:DeltaQ<fH_[m,m]}, namely $\Delta Q_{\mathrm{corr}\,A:B}\le f_H\!\bigl([m_{AM},m_{MB}]\bigr)$. Here, the increase of quantum correlations between the probes is upper-bounded by a function of the commutators of the mediator observables dynamically involved in the specific interaction.}

{
Taken together, these contributions establish a \emph{general-quantitative} witness of mediator non-classicality: by Corollary~\ref{cor:positive_gain_nonclassical_interactions}, observing $\Delta Q_{\mathrm{corr}\,A:B}>0$ rules out commuting local interactions; Result~\ref{res:nonclassical_interactions_imply_nonclassical_mediator} then ensures that at least one pair of mediator observables dynamically involved in the coupling is non-commuting, while the observed increase in quantum correlations is constrained by the bound \eqref{main_res_explicit} for all $t$.}

{
These results provide a general quantitative framework for the witness of non-classicality in \cite{PRL_Marletto_Vedral_2017,bose_spin_2017}.
In those proposals, the creation of entanglement between two probes interacting locally only through a mediator is a sufficient condition to infer the mediator's non-classicality.
Here, within the same mediated-interaction setting, we make this connection quantitative and independent of the particular choice of interaction Hamiltonians by relating the observed gain in quantum correlations to a function controlled by the non-commutativity of the mediator's observables.}

We further explored the impact of decoherence on the mediator's observables, particularly through {a specific type of} phase-flip channel. We showed that as the phase-flip parameter $p$ approaches $ \frac{1}{2} $, the {commutativity of mediator's observables} reaches zero, effectively yielding the mediator to be classical and preventing its ability to increase quantum correlations between two probes.

Future work can expand on our findings in several ways. One promising direction is to interpret the mediator as a quantum channel.
In particular, if one chooses as correlation measure $Q$ the relative entropy of entanglement $E_{A:B}$, it is natural to define the corresponding correlation gain $\Delta E_{A:B}$.
With this choice, our general bound \eqref{eq:DeltaQ<fH_[m,m]} reads
\begin{equation}
\Delta E_{A:B}\;\le\; f_H\!\bigl([m_{AM},m_{MB}]\bigr),
\label{eq:REE_bound_fH}
\end{equation}
where $f_H$ is fixed once $t$, and the Hamiltonians are specified, and it is controlled by the commutators of the mediator observables dynamically involved in the coupling.
A natural question is whether a gain in relative entropy of entanglement $\Delta E_{A:B}$ can be related to a quantum capacity or to the entanglement generation capacity of the mediator $M$, as suggested in \cite{bennett_capacities_2003}.
This involves investigating whether one can translate the two sides of \eqref{eq:REE_bound_fH} into bounds on channel capacities of $M$.
More precisely, one would like to establish quantitative relations of the schematic form
\begin{equation}
\mathcal{Q}^{(\min)}_{M}\!\bigl(\Delta E_{A:B}\bigr)\;\;\le\;\;
\mathcal{Q}^{(\max)}_{M}\!\Bigl(f_H\!\bigl([m_{AM},m_{MB}]\bigr)\Bigr),
\label{eq:Qmin_Qmax_from_REE_bound}
\end{equation}
where $\mathcal{Q}_{M}^{(\min)}$ denotes a chosen capacity of the mediator and is an experimentally inferred lower bound constructed from the observed entanglement gain, while $\mathcal{Q}^{(\max)}_{M}$ is a theoretical upper bound controlled by the mediator commutators activated by the interaction. Interestingly, this transition could be obtained by relating \eqref{eq:REE_bound_fH} in terms of coherent information, whose supremum is indeed the quantum channel capacity \cite{schumacher_quantum_1996,lloyd_capacity_1997}.

Within this context, two distinct situations arise: testing the non-classicality of an unknown system (e.g., a biological system) which is suspected to have quantum properties, by attempting quantum communication with it and measuring quantum correlation gain in the probes. Alternatively, utilising the mediator for entanglement distribution protocols across a quantum network, one can theoretically calculate a bound for its maximum ability to deliver entanglement between two probes without prior testing. In both cases, our inequality can provide bounds on the quantum capacity of the mediator based on its observable and Hilbert space structure. 

Additionally, in the context of super-selected models for gravity \cite{sherry_interaction_1978}, where the classicality of gravity emerges from imposing certain superselection rules to commutators of its quantum observables, our result can be used as a falsification test: if quantum correlations are detected between probes, our inequality \eqref{eq:DeltaQ<fH_[m,m]} implies that quantum gravitational observables $m_j, m_l$ are dynamically involved, potentially leading to the exclusion of certain semi-classical models for gravity.

Finally, the symmetry between two-point space and time quantum correlations \cite{zhao_geometry_2018} suggests the possibility of extending the results of this work to mediated temporal quantum correlation between a single quantum probe at times $t_0$ and $t>t_0$. The potential differences in the functional form of the upper bound \eqref{eq:DeltaQ<fH_[m,m]} can inform on the symmetrical role of space and time in quantum mechanics, and offer a common framework to treat the witness of non-classicality in \cite{bose_spin_2017, PRL_Marletto_Vedral_2017,marletto_witnessing_2020} and its temporal, single-probe, version, as discussed in \cite{di_pietra_temporal_2023,feng_conservation_2024, dipietraTemporalEntanglementWitnesses2025}.

In summary, our proposed inequality gives a \textit{general quantitative framework} for the recently proposed witnesses of non-classicality  \cite{PRL_Marletto_Vedral_2017,bose_spin_2017}. We have also demonstrated the quantum-to-classical transition of the mediator by studying a decohering map applied to the mediator.
Our work thus provides a fundamental link between the mediator's non-classical features and the observed quantum correlations, independent of the specific dynamics of the interactions and the choice of the type of systems used. Its generality makes it suitable for a range of possible applications, from quantum metrology to quantum biology and even quantum gravity. 
\\

\textbf{Acknowledgements} \;  
We thank Vlatko Vedral, Simone Rijavec, Tomasz Paterek {and one anonymous Referee} for helpful comments and discussions.
G.D.P. thanks the Clarendon Fund and the Oxford-Thatcher Graduate Scholarship for supporting this research.
C.M. thanks the Eutopia Foundation, the John Templeton Foundation, and the Gordon and Betty Moore Foundation. This publication was made possible through the support of the ID 61466 grant from the John Templeton Foundation, as part of The Quantum Information Structure of Spacetime (QISS) Project (qiss.fr). The opinions expressed in this publication are those of the Authors and do not necessarily reflect the views of the John Templeton Foundation.

\bibliographystyle{apsrev4-2}
\bibliography{bibliography}

\newpage
\appendix

\counterwithout{equation}{section} 

\setcounter{equation}{0}
\renewcommand{\thesection}{A.\Roman{section}}

\renewcommand{\thesubsection}{A.\Roman{subsection}}
\renewcommand\theequation{A.\arabic{equation}}

\section{Definitions and Notation}
\label{sec:appendixA1}
{
This section briefly revisits essential definitions and notation used throughout the paper, and introduces further notation required for the derivations presented later in this Appendix.\\}

{\noindent\textbf{System configuration.}
The system under investigation, considered as \textit{closed}, consists of two quantum probes, $A$ and $B$, and a mediator $M$ only. Each subsystem is described by a finite-dimensional Hilbert space, denoted $\mathcal{H}_A$, $\mathcal{H}_M$, and $\mathcal{H}_B$, with dimensions $d_A=\dim(\mathcal{H}_A)$, $d_M=\dim(\mathcal{H}_M)$, and $d_B=\dim(\mathcal{H}_B)$, respectively. To enforce locality, i.e.\ to prohibit direct interactions between $A$ and $B$, we assume that the system Hamiltonian has the mediated form
\begin{equation}
\hat{H}=\hat{H}_{AM}+\hat{H}_{MB}.
\label{eq:app_H_total}
\end{equation}
In particular, no direct $A$--$B$ coupling term is present.} {Notice that the Hamiltonians $H_{AM}$ and $H_{MB}$ also include the free evolution of $A$ and $B$.}

\medskip
{
\noindent\textbf{Unitaries, Zassenhaus factorisation, and $U_C$.}
The unitary evolution associated with \eqref{eq:app_H_total} is
\begin{equation}
U_{AMB}(t)\vcentcolon=e^{-i\hat{H}t}=e^{-i(\hat{H}_{AM}+\hat{H}_{MB})t}.
\label{eq:app_U_total_def}
\end{equation}
We also define the local evolutions
\begin{equation}
U_{AM}(t)\vcentcolon=e^{-i\hat{H}_{AM}t},
\qquad
U_{MB}(t)\vcentcolon=e^{-i\hat{H}_{MB}t}.
\label{eq:app_U_local_def}
\end{equation}
Using the Zassenhaus expansion, one can factorise $U_{AMB}(t)$ as
\begin{equation}
U_{AMB}(t)
=
U_{AM}(t)\,U_{MB}(t)\,
\prod_{n=2}^{\infty} \exp\!\bigl[(-it)^n Z_n\bigr],
\label{eq:app_zassenhaus_exact}
\end{equation}
where the Zassenhaus exponents $Z_n$ are uniquely determined by the pair $\{\hat H_{AM},\hat H_{MB}\}$.
More precisely, for each $n\ge2$, $Z_n$ is a homogeneous Lie polynomial of degree $n$ in the two generators
$\hat H_{AM}$ and $\hat H_{MB}$ \cite{casas_efficient_2012}, i.e.\ a finite linear combination (with rational coefficients)
of nested commutators built from $\{\hat H_{AM},\hat H_{MB}\}$:
\begin{equation}
[V_1,[V_2,\ldots,[V_{n-1},V_n]\ldots]] : V_i\in\{\hat H_{AM},\hat H_{MB}\}.
\label{eq:Zn_nested_comm_structure}
\end{equation}
In particular, the leading Zassenhaus exponent is
\begin{equation}
Z_2= -\frac{1}{2}\,[\hat{H}_{AM},\hat{H}_{MB}].
\label{eq:app_Z2}
\end{equation}}

{The unitary $U_C(t)$, introduced in the main text in \eqref{eq:Ualph_Halp_def_main}, is given by the Zassenhaus correction appearing in \eqref{eq:app_zassenhaus_exact}. We define
\begin{equation}
U_C(t)
\vcentcolon=
\prod_{n=2}^{\infty} \exp\!\bigl[(-it)^n Z_n\bigr]
=
U_{MB}^\dagger(t)\,U_{AM}^\dagger(t)\,U_{AMB}(t),
\label{eq:app_Ualpha_def}
\end{equation}
and also the corresponding Hamiltonian generator $\hat H_C$ implicitly by
\begin{equation}
U_C(t)=e^{-it\hat H_C}.
\label{eq:app_Halpha_def}
\end{equation}}

{Moreover, if the interaction Hamiltonians commute, then every nested commutator built from
$\{\hat H_{AM},\hat H_{MB}\}$, as introduced in \eqref{eq:Zn_nested_comm_structure}, vanishes. Since each $Z_n$ is a homogeneous Lie polynomial of degree $n\ge2$ in these two generators, this implies \begin{equation}
[\hat H_{AM},\hat H_{MB}]=0
\;\Longrightarrow\;
Z_n=0
\quad \forall\, n\ge 2,
\end{equation}
and therefore each factor in the Zassenhaus product is the identity,
\begin{equation}
\exp\!\bigl[(-it)^n Z_n\bigr]=\mathbb{I}
\quad \forall\, n\ge 2.
\label{eq:commuting_implies_each_factor_trivial}
\end{equation}
Hence, the entire Zassenhaus correction is trivial,
\begin{equation}
U_C(t)=\mathbb{I},
\qquad\text{and consequently}\qquad
\hat H_C=0.
\end{equation}}

{Independently, in the commuting case, the Baker--Campbell--Hausdorff formula yields the
exact factorisation of the global unitary
\begin{equation}
\begin{split}
U_{AMB}(t)
&= e^{-it(\hat H_{AM}+\hat H_{MB})}
= e^{-it\hat H_{AM}}\,e^{-it\hat H_{MB}} \\
&= U_{AM}(t)\,U_{MB}(t),
\qquad \text{for } [\hat H_{AM},\hat H_{MB}]=0.
\end{split}
\label{eq_app:commuting_exact_factorisation}
\end{equation}
By the definition \eqref{eq:app_Ualpha_def}, this exact factorisation gives
$U_C(t)=\mathbb{I}$ and hence $\hat H_C=0$, in agreement with the
Zassenhaus-based argument above.}

{We therefore summarise these implications as
\begin{equation}
[\hat H_{AM},\hat H_{MB}]=0
\;\Longrightarrow\;
U_C(t)=\mathbb{I}
 \;\Longrightarrow\;
\hat H_C=0.
\label{eq:commuting_implies_Ualpha_trivial}
\end{equation}
Equivalently, a nontrivial Zassenhaus correction implies non-commuting interactions,
namely
\begin{equation}
U_C(t)\neq\mathbb{I} \;\Longrightarrow\; \hat H_C\neq 0
\;\Longrightarrow\;
[\hat H_{AM},\hat H_{MB}]\neq 0.
\label{eq:Halpha_nonzero_implies_noncommuting}
\end{equation}}

\medskip
{
\noindent\textbf{States and induced map.}
Let $\rho_{AMB}(0)$ denote the initial global state. The reduced initial state of the probes is
\begin{equation}
\rho_0 \vcentcolon= \rho_{AB}(0)=\mathrm{tr}_M\!\big[\rho_{AMB}(0)\big].
\label{eq:app_rho0}
\end{equation}
The evolved global state is
\begin{equation}
\rho_{AMB}(t)=U_{AMB}(t)\,\rho_{AMB}(0)\,U_{AMB}^\dagger(t),
\label{eq:app_global_state}
\end{equation}
hence the reduced final state of the probes is
\begin{equation}
\rho_t \vcentcolon= \rho_{AB}(t)=\mathrm{tr}_M\!\big[\rho_{AMB}(t)\big].
\label{eq:app_rhot}
\end{equation}
The corresponding global unitary channel is
\begin{equation}
\Lambda_{AMB}(\rho)\vcentcolon=U_{AMB}(t)\,\rho\,U_{AMB}^\dagger(t).
\label{eq:app_global_channel}
\end{equation}
Accordingly, the induced (marginal) map on the probes, $\Lambda_{AB}$, is defined by
\begin{equation}
\Lambda_{AB}(\rho) = \mathrm{tr}_M[\Lambda_{AMB}(\rho)].
\label{eq:app_induced_map}
\end{equation}}

\medskip
{
\noindent\textbf{Decomposability and its variants.}
Decomposability formalises the idea that the mediated dynamics can be realised in a single sequential ``round'': first an operation involving $AM$, followed by an operation involving $MB$.
At the level of channels, a CPTP map $\lambda_{AMB}$ on the tripartite system is said to be \emph{decomposable} if there exist CPTP maps $\lambda_{AM}$ and $\lambda_{MB}$ such that
\begin{equation}
\lambda_{AMB}(\rho)=\lambda_{MB}\circ \lambda_{AM}(\rho).
\label{eq:app_decomp_def}
\end{equation}
The ordering in \eqref{eq:app_decomp_def} is part of the statement: decomposability asserts the existence of \emph{at least one} factorisation of this form, and it does not, in general, imply that the reverse order $\lambda_{AMB}=\lambda_{AM}\circ\lambda_{MB}$ is also available.}

{For closed dynamics, the corresponding notion is defined directly at the level of unitaries. A unitary $U_{AMB}$ is \emph{decomposable} if it can be written as
\begin{equation}
U_{AMB}=U_{MB}\,U_{AM},
\label{eq:app_unitary_decomp_def}
\end{equation}
for some unitaries $U_{AM}$ and $U_{MB}$. When $U_{AMB}$ is generated by a time-independent Hamiltonian, $U_{AMB}(t)=e^{-it\hat{H}}$, one may also distinguish the stronger property of a \emph{continuous commuting decomposition}, namely the existence of continuous families $U_{AM}(t)$ and $U_{MB}(t)$ such that
\begin{equation}
\begin{split}
&U_{AMB}(t)=U_{AM}(t)\,U_{MB}(t)=U_{MB}(t)\,U_{AM}(t), \\  &U_{AM}(t),U_{MB}(t)]=0\quad \forall t.
\label{eq:app_commuting_decomposition}
\end{split}
\end{equation}
This strengthening is relevant because, as shown in \cite{ganardi_quantitative_2024}, commuting decompositions capture precisely the sense in which a unitary evolution can be regarded as generated by commuting local Hamiltonians, whereas plain decomposability alone is strictly weaker.}

\medskip
{\noindent\textbf{From tripartite decomposability to inaccessible mediators.}
In our setting, only the marginal evolution of the probes is operationally accessible, namely the CPTP map $\Lambda_{AB}$, through measurements performed on $A$ and $B$ directly.
Accordingly, to assess whether $\Lambda_{AB}$ belongs to the decomposable class \eqref{eq:app_decomp_def}, we consider suitable tripartite dilations, introducing an explicit mediator of bounded dimension.
Following \cite{ganardi_quantitative_2024}, we say that $\Lambda_{AB}$ admits a \emph{decomposable $m$-dilation} if there exist a mediator system $M$ with $d_M\le m$, an initial state $\sigma_M$, and a decomposable channel $\lambda_{ABM}$ such that
\begin{equation}
\Lambda_{AB}(\rho)=\mathrm{tr}_M\!\big[\lambda_{ABM}(\rho\otimes\sigma_M)\big]
\qquad\forall\,\rho.
\label{eq:app_decomp_dilation_def}
\end{equation}
We denote by $\overline{DEC}(m)$ the set of all maps $\Lambda_{AB}$ admitting such a decomposable $m$-dilation.}

\medskip
{\noindent\textbf{Classicality of interactions and decomposability.}
For closed mediated dynamics generated by a local Hamiltonian of the type \eqref{eq:app_H_total}, a natural notion of mediated \emph{classical interaction} is the commutativity of the coupling terms,
\begin{equation}
[\hat{H}_{AM},\hat{H}_{MB}]=0.
\label{eq:app_commuting_H}
\end{equation}
In this case, the BCH formula implies an exact factorisation of the global unitary for all $t$, as already shown in \eqref{eq_app:commuting_exact_factorisation}. So the evolution is continuously decomposable with order-independent (commuting) factors in the sense of \eqref{eq:app_commuting_decomposition}.}

\medskip
{\noindent\textbf{Quantum correlation measure $Q$, correlation gain $\Delta Q_{\mathrm{corr}\,A:B}$, and the initial correlation $\mathcal{B}(Q(\rho_0))$}
Throughout this work, $Q$ denotes a generic bipartite quantifier of quantum correlations.
We assume that $Q$ belongs to the class of \emph{$gd$-continuous} functions, which follow a continuity property with
respect to a contractive distance $d$, namely there exists an invertible, monotonically
increasing function $g$, with $\lim_{s\to 0} g(s)=0$, such that for any pair of states $x$
and $y$ one has
\begin{equation}
|Q(x)-Q(y)| \leq g\!\bigl(d(x,y)\bigr).
\end{equation}
In addition, $Q$ is taken to be monotonic under local processing.}

{With $\rho_0$ and $\rho_t$ defined in \eqref{eq:app_rho0} and \eqref{eq:app_rhot}, we
quantify the correlation gain between $A$ and $B$ by
\begin{equation}
\Delta Q_{\mathrm{corr}\,A:B}
\vcentcolon=
Q_{A:B}(\rho_t) - \mathcal{B}\!\bigl(Q(\rho_0)\bigr).
\label{eq:DeltaQ_def_app}
\end{equation}
The term $Q_{A:B}(\rho_t)$ is directly accessible from measurements on the probes, whereas
$\mathcal{B}\!\bigl(Q(\rho_0)\bigr)$ accounts for the maximum potential correlation within the system before any interaction.
\\
The quantity $\mathcal{B}\!\bigl(Q(\rho_0)\bigr)$ is defined as
\begin{equation}
\mathcal{B}\!\bigl(Q(\rho_0)\bigr)
=
\sup_{\sigma_{AM}} Q_{A:M}(\sigma_{AM})
+
\sup_{\sigma_{MB}} Q_{M:B}(\sigma_{MB})
+
I_{A:B}(\rho_0),
\label{eq:app_B_def}
\end{equation}
where the supremum runs over all $AM$ and $MB$ states, while $I_{A:B}(\rho_0)$ is calculated as
\begin{equation}
I_{A:B}(\rho_0)
=
\inf_{\sigma_A \otimes \sigma_B}
g\!\left(d(\rho_{0}, \sigma_A \otimes \sigma_B)\right),
\label{eq:app_I_def}
\end{equation}
and the infimum runs over all product states $\sigma_A \otimes \sigma_B$. Here, $\sigma_A$
and $\sigma_B$ denote generic initial states of $A$ and $B$, respectively.}

\section{Connecting the mediated quantum correlations to the non-classicality of interactions}
\label{sec:appendixA2}

{In this section, we derive Result \ref{result1} and Eq. \eqref{main_res_explicit}, starting from the distance-based inequality used in \cite{ganardi_quantitative_2024}. The argument proceeds in two steps: first, we show that the quantity $g^{-1}\!\bigl(\Delta Q_{\mathrm{corr}\,A:B}\bigr)$ can be upper bounded by the spectral norm of a suitable difference of unitaries, $\|U-V\|_\infty$, associated with the global and local interaction terms of \eqref{eq:app_U_total_def} and \eqref{eq:app_U_local_def}. Second, we rewrite this unitary difference in terms of the Zassenhaus correction $U_C(t)$ and its generator $\hat H_C$, and we show that it reduces to the spectral norm of $U_C(t)-\mathbb{I}$, which can be evaluated with a closed functional form.}
\\

{We begin the derivation by recalling the distance-based inequality from \cite{ganardi_quantitative_2024}:
\begin{equation}
g^{-1}\!\left(\Delta Q_{\mathrm{corr}\,A:B}\right)
\le
d\!\Bigl(\widetilde{\Lambda}\!\bigl(\rho_{AB}(0)\otimes\sigma_M\bigr),
\widetilde{\lambda}\bigl(\rho_{AB}(0)\otimes\sigma_M\bigr)\Bigr).
\label{eq:C3_Ganardi_et_al_2024}
\end{equation}
It holds for arbitrary choices of the dilations $\widetilde{\Lambda},\widetilde{\lambda}$ of the marginal
map $\Lambda_{AB}$ and for all decomposable maps $\lambda_{AB}\in\overline{DEC}(m)$.
In particular, since the right-hand side provides an upper bound for any admissible dilation pair, we may take a convenient choice for $\widetilde{\Lambda}$ and for a decomposable $\widetilde{\lambda}$.
We pick the unitary dilation induced by the actual global evolution, and a decomposable unitary dilation, namely
\begin{equation}
\begin{split}
\widetilde{\Lambda}(\rho)
&\vcentcolon=
\Lambda_{AMB}(\rho)
=U_{AMB}\,\rho\,U_{AMB}^\dagger,\\
\widetilde{\lambda}(\rho)
&\vcentcolon=
\lambda_{AMB}(\rho)
=\bigl(U_{AM}U_{MB}\bigr)\,\rho\,\bigl(U_{MB}^\dagger U_{AM}^\dagger\bigr),
\end{split}
\label{eq:choice_of_dilations}
\end{equation}
where we keep the time argument $t$ implicit to lighten the notation. We also take $d$ to be the trace distance,
\begin{equation}
d_{\mathrm{tr}}(X,Y)\vcentcolon=\tfrac12\|X-Y\|_1.
\end{equation}
For compactness, define
\begin{equation}
U \vcentcolon= U_{AMB},
\quad
V \vcentcolon= U_{AM}U_{MB},
\quad
\rho \vcentcolon= \rho_{AB}(0)\otimes\sigma_M.
\label{eq:U_V_rho_notation}
\end{equation}
With these choices, \eqref{eq:C3_Ganardi_et_al_2024} becomes
\begin{equation}
\begin{split}
g^{-1}\!\bigl(\Delta Q_{\mathrm{corr}\,A:B}\bigr)
&\le
d_{\mathrm{tr}}\!\bigl(\Lambda_{AMB}(\rho),\,\lambda_{AMB}(\rho)\bigr)\\
&=
\frac{1}{2}\,\Bigl\|U\rho U^\dagger - V\rho V^\dagger\Bigr\|_1,
\end{split}
\label{eq:trace_dist_to_trace_norm}
\end{equation}
which we can rewrite as
\begin{equation}
\begin{split}
&\frac{1}{2}\, \Bigl\|U\rho U^\dagger - V\rho V^\dagger\Bigr\|_1
=
\frac12\Bigl\|(U-V)\rho U^\dagger + V\rho\,(U^\dagger - V^\dagger)\Bigr\|_1\\
&\le
\frac12\Bigl(\,\|(U-V)\rho U^\dagger\|_1+\|V\rho\,(U^\dagger - V^\dagger)\|_1\Bigr),
\end{split}
\label{eq:split_triangle}
\end{equation}
where we added and subtracted $V\rho U^\dagger$ and used the triangle inequality for $\|\cdot\|_1$. Next, using the ideal property of Schatten norms $\|AXB\|_1\le \|A\|_\infty\|X\|_1\|B\|_\infty$ \cite{schatten_norm_1960,watrous_theory_2018}, we obtain
\begin{equation}
\begin{split}
\frac12 &\Bigl(\,\|(U-V)\rho U^\dagger\|_1+\|V\rho\,(U^\dagger - V^\dagger)\|_1\Bigr)
\\\le
&\frac12\Bigl(\,\|U-V\|_\infty\,\|\rho\|_1\,\|U^\dagger\|_\infty
+\|V\|_\infty\,\|\rho\|_1\,\|U^\dagger - V^\dagger\|_\infty\Bigr).
\end{split}
\label{eq:holder_step}
\end{equation}
Finally, since $\|\rho\|_1=1$, and $U$ and $V$ are unitary so that $\|U\|_\infty=\|V\|_\infty=\|U^\dagger\|_\infty=1$, and
$\|U^\dagger - V^\dagger\|_\infty=\|U-V\|_\infty$, we conclude
\begin{equation}
g^{-1}\!\bigl(\Delta Q_{\mathrm{corr}\,A:B}\bigr)
\le
\|U-V\|_\infty.
\label{eq:trace_to_spectral}
\end{equation}}

{Let us now express the right-hand side of \eqref{eq:trace_to_spectral} in terms of the Zassenhaus correction
$U_C(t)$ introduced in Appendix \ref{sec:appendixA1}.
Using \eqref{eq:app_zassenhaus_exact} and \eqref{eq:app_Ualpha_def}, we have
\begin{equation}
U_{AMB}(t)=U_{AM}(t)\,U_{MB}(t)\,U_C(t).
\label{eq:UAMB_split_Ualpha}
\end{equation}
Therefore, given the definitions of $U$ and $V$ in \eqref{eq:U_V_rho_notation} we have
\begin{equation}
\begin{aligned}
& g^{-1}\!\bigl(\Delta Q_{\mathrm{corr}\,A:B}\bigr)
\le
\|U-V\|_\infty = \bigl\|U_{AMB}-U_{AM}U_{MB}\bigr\|_\infty
 \\
& = \bigl\|U_{AM}U_{MB}\bigl(U_C-\mathbb{I}\bigr)\bigr\|_\infty\le
\bigl\|U_{AM}U_{MB}\bigr\|_\infty\,
\bigl\|U_C-\mathbb{I}\bigr\|_\infty\\
&=
\bigl\|U_C-\mathbb{I}\bigr\|_\infty,
\end{aligned}
\label{eq:Udiff_to_Ualpha_minus_I}
\end{equation}
where we used submultiplicativity of $\|\cdot\|_\infty$ and the fact that products of unitaries have spectral norm equal to $1$. Thus, we obtain
\begin{equation}
g^{-1}\!\bigl(\Delta Q_{\mathrm{corr}\,A:B}\bigr)
\le
\bigl\|U_C-\mathbb{I}\bigr\|_\infty.
\label{eq:DeltaQ_to_Ualpha_minus_I}
\end{equation}
To manipulate the right-hand side, we recall the spectral norm definition
\begin{equation}
\|X\|_\infty \vcentcolon=
\sqrt{\lambda_{\max}\!\bigl(X^\dagger X\bigr)}.
\label{eq:infty_norm_def}
\end{equation}
Evaluating the square of \eqref{eq:infty_norm_def} with $X=U_C(t)-\mathbb{I}$, yields
\begin{equation}
\begin{aligned}
\bigl\|U_C-\mathbb{I}\bigr\|_\infty^2
&=
\lambda_{\max}\!\Bigl(
\bigl(U_C-\mathbb{I}\bigr)^\dagger
\bigl(U_C-\mathbb{I}\bigr)
\Bigr)\\
&=
\lambda_{\max}\!\Bigl(
\bigl(e^{it\hat{H}_C}-\mathbb{I}\bigr)
\bigl(e^{-it\hat{H}_C}-\mathbb{I}\bigr)
\Bigr)\\
&=
\lambda_{\max}\!\Bigl(
2\mathbb{I}-e^{it\hat{H}_C}-e^{-it\hat{H}_C}
\Bigr)\\
&=
\lambda_{\max}\!\Bigl(
2\bigl(\mathbb{I}-\cos(t\hat{H}_C\bigr)
\Bigr),
\end{aligned}
\label{eq:Ualpha_cos_step}
\end{equation}
where we used the definition \eqref{eq:app_Halpha_def} for $U_C$ and the operatorial identity $e^{iY}+e^{-iY}=2\cos(Y)$ for a Hermitian $Y$.
Using the trigonometric identity $1-\cos(Y)=2\sin^2(Y/2)$ and taking square roots gives
\begin{equation}
\bigl\|U_C-\mathbb{I}\bigr\|_\infty
=
2\,\Bigl\|\sin\!\Bigl(\frac{t}{2}\hat{H}_C\Bigr)\Bigr\|_\infty,
\label{eq:Ualpha_minus_I_sine}
\end{equation}
and therefore we find the Result \ref{result1} given in \eqref{main_res_explicit}
\begin{equation}
g^{-1}\!\bigl(\Delta Q_{\mathrm{corr}\,A:B}\bigr)
\le
2\,\Bigl\|\sin\!\Bigl(\frac{t}{2}\hat{H}_C\Bigr)\Bigr\|_\infty.
\label{eq:DeltaQ_to_Halpha_sine}
\end{equation}}

{The closed form \eqref{eq:DeltaQ_to_Halpha_sine} also allows one to characterise the maximal value of the bound and the earliest interaction time at which it can be reached.
Since $\|\sin(X)\|_\infty\le 1$ for any Hermitian $X$, the right-hand side of \eqref{eq:DeltaQ_to_Halpha_sine} is bounded by $2$.
Let $\{\lambda_j\}$ denote the eigenvalues of $\hat H_C$, then we have
\begin{equation}
\Bigl\|\sin\!\Bigl(\frac{t}{2}\hat H_C\Bigr)\Bigr\|_\infty
=
\max_j \Bigl|\sin\!\Bigl(\frac{t}{2}\lambda_j\Bigr)\Bigr|.
\label{eq:spectral_mapping_sin}
\end{equation}
Fix $j$ with $\lambda_j\neq 0$. The modulus $\bigl|\sin(\tfrac{t}{2}\lambda_j)\bigr|$ attains its maximum value~$1$
if and only if
\begin{equation}
\frac{t}{2}\lambda_j=\frac{\pi}{2}+k\pi,
\qquad k\in\mathbb{Z},
\end{equation}
since $\sin(\frac{\pi}{2}+k\pi)=(-1)^k$.
Accordingly, the set of interaction times for which the $j$-th eigenvalue saturates the modulus is
\begin{equation}
t_{j,k}=\frac{\pi+2k\pi}{\lambda_j},
\qquad k\in\mathbb{Z},
\end{equation}
and the corresponding \emph{smallest positive} time is obtained by taking $k=0$ and using the modulus,
\begin{equation}
t_j^\ast = \frac{\pi}{|\lambda_j|}.
\end{equation}
Therefore, the earliest time at which the operator norm can reach its maximum is
\begin{equation}
t_\ast \vcentcolon= \min_j t_j^\ast = \min_j \frac{\pi}{|\lambda_j|}
= \frac{\pi}{\max_j |\lambda_j|}.
\label{eq:tstar_def}
\end{equation}
Let $\lambda_{\max}$ be an eigenvalue such that $|\lambda_{\max}|=\max_j|\lambda_j|$. Then at $t=t_\ast$ we have
\begin{equation}
\biggl|\sin\Big(\frac{t_\ast}{2}\lambda_{\max}\Big)\biggr|=1, \quad \text{and hence} \quad \bigg|\sin\Big(\frac{t_\ast}{2}\hat H_C\Big)\bigg|_\infty=1.    
\end{equation}
In particular, for $\hat H_C\neq 0$, the right-hand side of \eqref{eq:DeltaQ_to_Halpha_sine} attains its maximal value~$2$ at $t=t_\ast$.}

{Finally, although \eqref{eq:DeltaQ_to_Halpha_sine} holds exactly for all interaction times \(t\), it is useful to extract a more explicit commutator-based expression in the short-time regime. In particular, truncating \(U_C\), as defined in \eqref{eq:app_Ualpha_def}, to the leading nontrivial Zassenhaus term \((n=2)\), yields
\begin{equation}
U_C(t)=\exp\!\bigl[(-it)^2 Z_2\bigr].
\end{equation}}
{Using the explicit expression for \(Z_2\) given in \eqref{eq:app_Z2}, we then obtain}
 
{\begin{equation}
U_C(t)=
\exp\!\Bigl(-it\,\hat{H}_C(t)\Bigr) =  \exp\!\Bigl(\tfrac{t^2}{2}\,[\hat{H}_{AM}, \hat{H}_{MB}]\Bigr),
\label{eq:Ualpha_Z2_approx}
\end{equation}
which implies
\begin{equation}
\hat{H}_C = \frac{i t}{2}\,[\hat{H}_{AM}, \hat{H}_{MB}].
\label{eq:app_Halpha_Z2_approx}
\end{equation}
Substituting this approximation into \eqref{eq:DeltaQ_to_Halpha_sine} gives the explicit commutator form
\begin{equation}
g^{-1}\!\bigl(\Delta Q_{\mathrm{corr}\,A:B}\bigr)
\leq
2\,\Bigl\|\sin\!\Bigl(\frac{i t^2}{4}\,[\hat{H}_{AM}, \hat{H}_{MB}]\Bigr)\Bigr\|_\infty,
\label{eq:DeltaQ_commutator_bound}
\end{equation}
which matches the small-$t$ version \eqref{main_res_explicit_[]}.}

\section{Non-classicality of interactions depends on the non-classicality of the mediator}\label{sec:appendixA3}
{In this Appendix section, we shall show how the commutator between the local interaction Hamiltonians $\hat{H}_{AM}, \hat{H}_{MB}$ in a tripartite system $AMB$, depends on the non-classicality of the investigated system $M$. We recall that we define a system as \textit{non-classical} when its description requires a minimum of two non-commuting observables, i.e., $m_{j_1},m_{j_2}$ such that $[m_{j_1},m_{j_2}]\neq0, {j_1}\neq {j_2}$}. 

{We begin by writing the most general form of the local interaction Hamiltonians $\hat H_{AM}$ and $\hat H_{MB}$ used in the main text \eqref{eq:most_general_HAM_HMB}, which we reproduce here for convenience:}
\begin{equation}
\begin{split}
&\hat{H}_{AM} = \\
& = \sum_{i, j_1, j_2, \ldots j_{T}}\alpha_{i, j_1, j_2, \ldots j_{T}} (a_i \otimes m_{j_1}\otimes m_{j_2} \otimes \ldots \otimes  m_{j_{T}} \otimes \mathbb{I}_{B}) \\
& \\
&\hat{H}_{MB} = \\
& =\sum_{l_1, l_2, \ldots , l_{T}, k}\beta_{l_1, l_2, \ldots , l_{T}, k} (\mathbb{I}_{A} \otimes m_{l_1}\otimes m_{l_2} \otimes \ldots \otimes  m_{l_{T}} \otimes b_k){.}
\label{eq:app_HAM_HMB_def}
\end{split}
\end{equation}
We can now write the commutator
\begin{equation}\begin{split}
&[\hat{H}_{AM}, \hat{H}_{MB}] = \\
&\Bigg[\sum_{i, j_1, j_2, \ldots j_{T}}\alpha_{i, j_1, j_2, \ldots j_{T}} (a_i \otimes m_{j_1}\otimes m_{j_2} \otimes \ldots \otimes  m_{j_{T}} \otimes \mathbb{I}_{B}), \\
&\sum_{l_1, l_2, \ldots , l_{T}, k}\beta_{l_1, l_2, \ldots , l_{T}, k} (\mathbb{I}_{A} \otimes m_{l_1}\otimes m_{l_2} \otimes \ldots \otimes  m_{l_{T}} \otimes b_k)\Bigg].
\end{split}
\end{equation}
To simplify the notation, we factorise the sums and the coefficients out of the commutator by using its linearity.
\begin{equation}
\begin{split}
[\hat{H}_{AM}, \hat{H}_{MB}] = & \sum_{i, j_1, j_2, \ldots j_{T}, l_1, l_2, \ldots , l_{T}, k} \alpha_{i, j_1, j_2, \ldots j_{T}} \beta_{l_1, l_2, \ldots , l_{T}, k} \\ \bigg[
& a_i \otimes m_{j_1} \otimes m_{j_2} \otimes \ldots \otimes m_{j_{T}} \otimes \mathbb{I}_{B}, \\
& \mathbb{I}_{A} \otimes m_{l_1} \otimes m_{l_2} \otimes \ldots \otimes m_{l_{T}} \otimes b_k \bigg].
\end{split}
\end{equation}
Using the linearity of the commutator and distributivity of the tensor product again, we can factorise probe operators out as well.
\begin{equation}
\begin{split}
&[\hat{H}_{AM}, \hat{H}_{MB}] = \sum_{i, j_1, j_2, \ldots j_{T}, l_1, l_2, \ldots , l_{T}, k} \alpha_{i, j_1, j_2, \ldots j_{T}} \beta_{l_1, l_2, \ldots , l_{T}, k} \\
& a_i \otimes \left[ m_{j_1}\otimes m_{j_2} \otimes \ldots \otimes  m_{j_{T}}, m_{l_1}\otimes m_{l_2} \otimes \ldots \otimes  m_{l_{T}} \right] \otimes b_k.
\end{split}
\end{equation}
Now we can only focus on the commutator of mediator observables. We are also going to use the following property of the commutator
\begin{equation}
[A \otimes B, C\otimes D] = CA \otimes [B, D] + [A, C] \otimes BD,
\label{eq:app_commutator_formula}
\end{equation}
where $A, C \in \mathcal{H}$, $B, D \in \mathcal{H}'$ {with $\mathcal{H}$ and
$\mathcal{H}'$ denoting two different Hilbert spaces}. Then we set $ A = m_{j_1} , C = m_{l_1} $ and $ B = m_{j_2} \otimes \ldots \otimes  m_{j_{T}}, D = m_{l_2} \otimes \ldots \otimes  m_{l_{T}}$ obtaining
\begin{equation}
\begin{split}
&[m_{j_1} \otimes m_{j_2} \otimes \ldots \otimes  m_{j_{T}}, m_{l_1} \otimes m_{l_2} \otimes \ldots \otimes  m_{l_{T}}]  \\
& = \   m_{l_1} m_{j_1} \otimes [m_{j_2} \otimes \ldots \otimes  m_{j_{T}}, m_{l_2} \otimes \ldots \otimes  m_{l_{T}}] \\
&\quad + [m_{j_1}, m_{l_1}] \otimes m_{j_2} m_{l_2} \otimes \ldots \otimes  m_{l_{T}}.
\end{split}
\end{equation}
Now we can recursively apply the same formula {\eqref{eq:app_commutator_formula}} to the commutator in the first term of the sum in the previous equation, coming to the final result for the commutator of the Hamiltonians
\begin{equation}
\begin{split}
&[\hat{H}_{AM}, \hat{H}_{MB}] =\\
& = \sum_{i, j_1, j_2, \ldots j_{T}, l_1, l_2, \ldots , l_{T}, k} \alpha_{i, j_1, j_2, \ldots j_{T}}^{\cancel{'}} \beta_{l_1, l_2, \ldots , l_{T}, k}^{\cancel{'}} \quad a_i \otimes\\
& \quad \otimes \bigg( [m_{j_1}, m_{l_1}] \otimes m_{j_2}m_{l_2} \otimes \ldots \otimes  m_{j_{T}} m_{l_{T}}  \\
& \quad  \quad + \, m_{l_1} m_{j_1} \otimes [m_{j_2}, m_{l_2}] \otimes \ldots \otimes  m_{j_{T}} m_{l_{T}} \\
& \quad \quad + \, m_{l_1} m_{j_1} \otimes m_{l_2} m_{j_2} \otimes \ldots \otimes  [m_{j_{T}}, m_{l_{T}}]\bigg) \otimes b_k.
\label{eq:app_general[H_am,H_mb]([m1,m2])}
\end{split}
\end{equation}
{When considering just two qubits for the mediator, the commutator of the two Hamiltonians reduces to
\begin{equation}
\begin{split}
&[\hat{H}_{AM}, \hat{H}_{MB}] =  \sum_{i, j_1, j_2, l_1, l_2, k} \alpha_{i, j_1, j_2} \beta_{l_1, l_2, k} \quad a_i \otimes \\ \bigg(
& [m_{j_1}, m_{l_1}] \otimes m_{j_2} m_{l_2}  + m_{l_1} m_{j_1} \otimes [m_{j_2}, m_{l_2}] \bigg) \otimes b_k.
\end{split}
\label{eq:app_nonclassmediator}
\end{equation}}

{To determine when the two-qubit interaction Hamiltonians commute or not, it is sufficient to analyse the
mediator-dependent factor appearing in \eqref{eq:app_nonclassmediator}:
\begin{equation}
[ m_{j_1}, m_{l_1} ] \otimes m_{j_2} m_{l_2} + m_{l_1} m_{j_1} \otimes \left[ m_{j_2}, m_{l_2} \right].
\label{eq:app_original_[m,m] t.p mm + mm t.p. [m,m]}
\end{equation}
Indeed, the probe operators $a_i$ and $b_k$ factorise, so that $[\hat H_{AM},\hat H_{MB}] \neq 0$ only if the factor \eqref{eq:app_original_[m,m] t.p mm + mm t.p. [m,m]} is non-vanishing. We therefore simplify the combination of commutators and products of mediators'
observables into a more compact form.}

\begin{equation}
\begin{split}
&[ m_{j_1}, m_{l_1} ] \otimes m_{j_2} m_{l_2} + m_{l_1} m_{j_1} \otimes \left[ m_{j_2}, m_{l_2} \right] \\
&= \left( m_{j_1} m_{l_1} - m_{l_1} m_{j_1} \right) \otimes m_{j_2} m_{l_2}  \\ 
& \quad + \, m_{l_1} m_{j_1} \otimes \left( m_{j_2} m_{l_2} - m_{l_2} m_{j_2} \right)  \\
&= m_{j_1} m_{l_1} \otimes m_{j_2} m_{l_2} - m_{l_1} m_{j_1} \otimes m_{j_2} m_{l_2}  \\
& \quad + \, m_{l_1} m_{j_1} \otimes m_{j_2} m_{l_2} - m_{l_1} m_{j_1} \otimes m_{l_2} m_{j_2}  \\
&= m_{j_1} m_{l_1} \otimes m_{j_2} m_{l_2} - m_{l_1} m_{j_1} \otimes m_{l_2} m_{j_2}. \\
\label{eq:twoqubit_tensor_comm_step1}
\end{split}
\end{equation}
{In the first two lines, we expand the commutators and distribute the tensor product over sums. After multiplying out, the second and third terms cancel.
Next, we use the Pauli-algebra anti-commutation relations:}
\begin{equation}
m_{l_1}m_{j_1}=2\,\delta_{l_1 j_1}\,\mathbb{I}-m_{j_1}m_{l_1}.
\label{eq:pauli_swap_identity}
\end{equation}
{Substituting \eqref{eq:pauli_swap_identity} into \eqref{eq:twoqubit_tensor_comm_step1} and collecting terms gives}
\begin{equation}
\begin{split}
&= m_{j_1} m_{l_1} \otimes m_{j_2} m_{l_2} - (2 \delta_{l_1 j_1} \mathbb{I} - m_{j_1} m_{l_1}) \otimes m_{l_2} m_{j_2} \\
&= m_{j_1} m_{l_1} \otimes \left( m_{j_2} m_{l_2} + m_{l_2} m_{j_2} \right) - 2 \delta_{l_1 j_1} \mathbb{I} \otimes m_{l_2} m_{j_2}. \\
\end{split}
\end{equation}
Combining the previous steps yields
\begin{equation}
\begin{split}
& [ m_{j_1}, m_{l_1} ] \otimes m_{j_2} m_{l_2} + m_{l_1} m_{j_1} \otimes \left[ m_{j_2}, m_{l_2} \right] = \ldots =\\
& = m_{j_1} m_{l_1} \otimes \left\{ m_{j_2}, m_{l_2} \right\} - 2 \delta_{l_1 j_1} \mathbb{I} \otimes m_{l_2} m_{j_2}.
\label{[m,m] t.p mm + mm t.p. [m,m]}
\end{split}
\end{equation}
{Having obtained \eqref{[m,m] t.p mm + mm t.p. [m,m]}, we can now characterise when this mediator factor \eqref{eq:app_original_[m,m] t.p mm + mm t.p. [m,m]} vanishes, and hence when $[\hat H_{AM},\hat H_{MB}] = 0$. This is conveniently done by distinguishing whether the first-qubit indices satisfy $l_1\neq j_1$ or $l_1=j_1$ (an equivalent split could be performed using the second mediator qubit).}
\\

\noindent\textbf{Case analysis for the two-qubit mediator factor.}
{Starting from \eqref{[m,m] t.p mm + mm t.p. [m,m]}, we distinguish two cases.}

\begin{description} 
\item[1. $l_1 \neq j_1$] this condition implies $\delta_{l_1 j_1} = 0$, leaving $m_{j_1} m_{l_1} \otimes \left\{ m_{j_2}, m_{l_2} \right\}$ to be the only term that may be non-vanishing in \eqref{[m,m] t.p mm + mm t.p. [m,m]}. This is the case when: $$ 
\left\{ \begin{array}{l} {m_{l_2} = m_{j_2} = \mathbb{I}} \\ m_{l_2} = m_{j_2} = X \\ m_{l_2} = m_{j_2} = Y \\ m_{l_2} = m_{j_2} = Z \\ 
\end{array} \right. \text{or} \left\{ 
\begin{array}{l} m_{l_2} = \mathbb{I}, \ m_{j_2} = X, Y, Z \\ m_{l_2} = X, Y, Z, \ m_{j_2} = \mathbb{I} \\ \end{array} \right. $$ 
{As a result, the second commutator appearing in \eqref{eq:app_original_[m,m] t.p mm + mm t.p. [m,m]} vanishes, and this yields to 
\begin{equation} 
\begin{split} &[ m_{j_1}, m_{l_1} ] \otimes m_{j_2} m_{l_2} + m_{l_1} m_{j_1} \otimes \left[ m_{j_2}, m_{l_2} \right] =\\ & = [ m_{j_1}, m_{l_1} ] \otimes m_{j_2} m_{l_2} .
\end{split} 
\end{equation} 
{Thus, in the case $l_1\neq j_1$, the non-vanishing contribution to the mediator factor can be traced back to
non-commutativity of the first-qubit operators, i.e.\ to $[m_{j_1},m_{l_1}]\neq 0$.} } \item[2. $l_1 = j_1$] after some simple manipulations \eqref{eq:app_original_[m,m] t.p mm + mm t.p. [m,m]} reads: 
\begin{equation} 
\begin{split} &[ m_{j_1}, m_{l_1} ] \otimes m_{j_2} m_{l_2} + m_{l_1} m_{j_1} \otimes \left[ m_{j_2}, m_{l_2} \right] =\\ & = \mathbb{I} \ \otimes \left[ m_{j_2}, m_{l_2} \right].\\ \end{split} 
\end{equation} 
Therefore, in this second case, when $l_1=j_1$, the mediator factor \eqref{eq:app_original_[m,m] t.p mm + mm t.p. [m,m]} is
non-vanishing only if the second-qubit contributes with non-commuting observables, i.e.\ only if $[m_{j_2},m_{l_2}]\neq 0$. \end{description} 

\noindent
In summary, in the two-qubit expansion \eqref{eq:app_nonclassmediator}, the commutator
$[\hat H_{AM},\hat H_{MB}]$ can be nonzero only if at least one pair of mediator Pauli operators, appearing in the two Hamiltonians, are non-commuting (either on the first or on the second mediator qubit).

\section{Dephasing mediator's qubits in Heisenberg Picture}\label{sec:appendixA4}
In this fourth Appendix section, we will rewrite the Hamiltonians defined in \eqref{eq:app_HAM_HMB_def} using the descriptors formalism {\cite{deutsch_information_2000,bedard_abc_2021}}, and recalculate their commutator after applying the dephasing channel to each of the mediator's qubits.

{Explicitly, the two interaction Hamiltonians can be written in the following form:}
\begin{equation}
\hat{H}_{AM} = \sum_{a_i, j_1, \ldots, j_T} \alpha_{a_i, j_1, \ldots, j_T} (q_{a_i} q_{j_1} \cdot \ldots \cdot q_{j_T} q_{\mathbb{I}_{B}}),
\label{eq:app_HAM_descriptors}
\end{equation}
\begin{equation}
\begin{split}
\hat{H}_{MB} &= \sum_{l_1, \ldots, l_T, b_k} \beta_{l_1, \ldots, l_T, b_k} (q_{\mathbb{I}_{A}} q_{l_1} \cdot \ldots \cdot q_{l_T}  q_{b_k}).
\end{split}
\end{equation}
As explained in Section \ref{sec:3}, the tensor product is substituted by a standard product since all the descriptors belong to $AMB$ Hilbert space. To preserve the locality condition, the identity operators of the original Hamiltonians \eqref{eq:app_HAM_HMB_def} are substituted by the descriptors $q_{\mathbb{I}_{B}}$ and $q_{\mathbb{I}_{A}}$, which are a tensor product of identity operators of all the qubits in the system.

Let us now apply a dephasing channel, as defined in \eqref{eq:Kraus_dephasing}, to the first qubit of the mediator on the Hamiltonian $\hat{H}_{AM}$ expressed in terms of descriptors \eqref{eq:app_HAM_descriptors}
\begin{equation}
\begin{split}
&\hat{E}_{1}(\hat{H}_{AM})  \equiv \sum_{a} M_{a}(t)^{\dagger} \hat{H}_{AM} \ M_{a}(t) = \\
& = \sqrt{p} \ I_{M_1}\hat{H}_{AM}I_{M_1} \sqrt{p} + \sqrt{1 - p} \ q_{z_{1}}^{\dagger}(t) \ \hat{H}_{AM} \ q_{z_{1}}(t) \sqrt{1 - p} \\
& = p \sum_{a_i, j_1, \ldots, j_T} \alpha_{a_i, j_1, \ldots, j_T} (q_{a_i}  I_{M_1} q_{j_1} I_{M_1}  \cdot \ldots \cdot q_{j_T} q_{\mathbb{I}_{B}}) \\
& \quad + (1 - p) \sum_{a_i, j_1, \ldots, j_T} \alpha_{a_i, j_1, \ldots, j_T}  ( q_{a_i}  q_{z_{1}} q_{j_1}q_{z_{1}}  \cdot \ldots \cdot  q_{j_T}  q_{\mathbb{I}_{B}}).  \\
\end{split}
\end{equation}
We now expand the second sum on the index \( j_1 \) and use the algebraic rule of the descriptors to compute $q_{z_{1}} q_{j_1}q_{z_{1}}$.
\begin{equation}
\begin{split}
&= p \sum_{a_i, j_1, \ldots, j_T} \alpha_{a_i, j_1, \ldots, j_T} (q_{a_i}  q_{j_1}  \cdot \ldots \cdot  q_{j_T}  q_{\mathbb{I}_{B}}) + (1-p)\\
& \quad \Bigg( \sum_{a_i, j_2, \ldots, j_T} \alpha_{a_i, j_1 = \mathbb{I}_{1}, j_2, \ldots, j_T}  (q_{a_i}  q_{\mathbb{I}_{1}}   \cdot \ldots \cdot  q_{j_T}  q_{\mathbb{I}_{B}})  \\
& \quad + \sum_{a_i, j_2, \ldots, j_T} \alpha_{a_i, j_1 = z_1, j_2, \ldots, j_T}  (q_{a_i}  q_{z_{1}}   \cdot \ldots \cdot  q_{j_T}  q_{\mathbb{I}_{B}})  \\
& \quad - \sum_{a_i, j_2, \ldots, j_T} \alpha_{a_i, j_1 = x_1, j_2, \ldots, j_T} (q_{a_i}  q_{x_{1}}  \cdot \ldots \cdot  q_{j_T}  q_{\mathbb{I}_{B}})  \\
& \quad - \sum_{a_i, j_2, \ldots, j_T} \alpha_{a_i, j_1 = y_1, j_2, \ldots, j_T}  q_{a_i}  q_{y_{1}}  \cdot \ldots \cdot ( q_{j_T}  q_{\mathbb{I}_{B}}) \Bigg) .\\
\end{split}
\end{equation}
Now, we expand the first sum on the index \( j_1 \) and sum the equivalent terms
\begin{equation}
\begin{split}
& =  \sum_{a_i, j_2, \ldots, j_T} \alpha_{a_i, j_1 = \mathbb{I}_{1}, j_2, \ldots, j_T}  (q_{a_i}  q_{\mathbb{I}_{1}}   \cdot \ldots \cdot  q_{j_T}  q_{\mathbb{I}_{B}}) \quad  \\
& + \sum_{a_i, j_2, \ldots, j_T} \alpha_{a_i, j_1 = z_1, j_2, \ldots, j_T} (a_{i}  q_{z_{1}} \cdot \ldots \cdot  q_{j_T}  q_{\mathbb{I}_{B}}) \quad \\
& + (2p- 1) \sum_{a_i, j_2, \ldots, j_T} \alpha_{a_i, j_1 = x_1, j_2, \ldots, j_T} (q_{a_i}  q_{x_{1}}    \cdot \ldots \cdot  q_{j_T}  q_{\mathbb{I}_{B}})  \\
& + (2p- 1) \sum_{a_i, j_2, \ldots, j_T} \alpha_{a_i, j_1 = y_1, j_2, \ldots, j_T} (q_{a_i}  q_{y_{1}}  \cdot \ldots \cdot  q_{j_T}  q_{\mathbb{I}_{B}})  . \\   
\end{split}
\end{equation}
Eventually, we absorb the four summation symbols and redefine a new coefficient $\alpha^{'}$, which takes into account a factor \((2p - 1)\) when multiplied by \( q_{x} \) or \( q_{y} \). This new primed coefficient now considers the effect of the phase flip on the first qubit of the mediator.
\begin{equation}
\begin{split} 
& {\hat{E}_{1}(\hat H _{AM})}= \sum_{a_i, j_1, \ldots, j_T} \alpha_{a_i, j_1, \ldots, j_T}^{'} (q_{a_i}  q_{j_1}  \cdot \ldots \cdot  q_{j_T}  q_{\mathbb{I}_{B}}).
\label{eq:app_degraded_ge_hamiltonian_HAM}
\end{split}
\end{equation}
Further phase flip operators \( \hat{E_2}, \hat{E_3}, \ldots, \hat{E_T},  \) behave  equivalently to \( \hat{E_1}  \), modifying coefficients when the respective observables \( q_x \) and \( q_y \) appear, with an extra \((2p -1)\). 
Having derived the dephased Hamiltonians, we can proceed {to derive} their commutator, which appears in \eqref{Degraded_Hamiltonian_Commutator}. This is a straightforward task, as we can substitute back the tensor product of the observables $a_i \otimes m_{j_1}\otimes m_{j_2} \otimes \ldots \otimes  m_{j_{T}} \otimes \mathbb{I}_{B}$ and perform the same derivation of Appendix \ref{sec:appendixA3} which led to \eqref{eq:app_general[H_am,H_mb]([m1,m2])}, obtaining the final result:
\begin{equation}
\begin{split}
&[\hat{H}_{AM}, \hat{H}_{MB}] =\\
& = \sum_{a_i, j_1, j_2, \ldots j_{T}, l_1, l_2, \ldots , l_{T}, b_k} \alpha_{i, j_1, j_2, \ldots j_{T}}^{'} \beta_{l_1, l_2, \ldots , l_{T}, k}^{'} \quad a_i \\
& \otimes \bigg( [m_{j_1}, m_{l_1}] \otimes m_{j_2}m_{l_2} \otimes \ldots \otimes  m_{j_{T}} m_{l_{T}}  \\
& \quad \quad m_{l_1} m_{j_1} \otimes [m_{j_2}, m_{l_2}] \otimes \ldots \otimes  m_{j_{T}} m_{l_{T}}  \\
& \quad \quad m_{l_1} m_{j_1} \otimes m_{l_2} m_{j_2} \otimes \ldots \otimes  [m_{j_{T}}, m_{l_{T}}]\bigg) \otimes b_k,
\end{split}
\end{equation}
which can be re-expressed using the descriptors formalism as
\begin{equation}
\begin{split}
&[\hat{H}_{AM}, \hat{H}_{MB}] =\\
& = \sum_{a_i, j_1, j_2, \ldots j_{T}, l_1, l_2, \ldots , l_{T}, b_k} \alpha_{a_i, j_1, j_2, \ldots j_{T}}^{'} \beta_{l_1, l_2, \ldots , l_{T}, b_k}^{'} \quad q_{a_i} \\
&  \quad \bigg( [q_{j_1}, q_{l_1}]  q_{j_2}q_{l_2} \cdot \ldots \cdot q_{j_{T}} q_{l_{T}} + q_{l_1} q_{j_1} [q_{j_2}, q_{l_2}]  \cdot \ldots \cdot  q_{j_{T}} q_{l_{T}}  \\
& \quad +  q_{l_1} q_{j_1} q_{l_2} q_{j_2} \cdot \ldots \cdot   [q_{j_{T}}, q_{l_{T}}]\bigg)  q_{b_k}.
\end{split}
\end{equation}

\end{document}